\documentclass[aps,prb,twocolumn,groupedaddress,amsmath,amssymb,showpacs]{revtex4-1}
\usepackage{graphicx,graphics,color,epsfig}
\usepackage{amsmath}
\usepackage{amssymb}
\usepackage{amsfonts}
\usepackage{amsfonts}
\usepackage{epstopdf}
\usepackage{bm}
\usepackage{times,xspace}
\usepackage{color}
\usepackage{array}
\usepackage{multirow}
\DeclareGraphicsExtensions{.pdf,.png,.jpg}

\begin{document}


\title{Coexistence of non-Fermi liquid and Fermi liquid self-energies at all dopings in cuprates}

\author{Sujay Ray and Tanmoy Das}
\affiliation{Department of Physics, Indian Institute of Science, Bangalore, 560012, India.}

\date{\today}

\begin{abstract}
Various angle-dependent measurements in hole-doped cuprates suggested that Non-Fermi liquid (NFL) and Fermi-liquid (FL) self-energies coexist in the Brillouin zone. Moreover, it is also found that NFL self-energies survive up to the overdoped region where the resistivity features a global FL-behavior. To address this problem, here we compute the {\it momentum} dependent self-energy from a single band Hubbard model. The self-energy is calculated self-consistently by using a momentum-dependent density-fluctuation (MRDF) method. One of our main result is that the computed self-energy exhibits a NFL-like frequency dependence only in the antinodal region, and FL-like behavior elsewhere, and retains its analytic form at all momenta and dopings. The dominant source of NFL self-energy in the antinodal region stems from the self-energy-dressed fluctuations between the itinerant and localized densities as self-consistency is invoked. We also calculate the DC conductivity by including the full momentum dependent self-energy. We find that the resistivity-temperature exponent $n$ becomes 1 near the optimal doping, while the NFL self-energy occupies largest momentum-space volume. Surprisingly, even in the NFL state near the optimal doping, the nodal region contains FL-like self-energies; while in the under- and over-dopings ($n\sim 2$), the antinodal region remains NFL-like. These results highlight the non-local correlation physics in cuprates and in other similar intermediately correlated materials, where a direct link between the microscopic single-particle spectral properties and the macroscopic transport behavior can not be well established. 
\end{abstract}

\pacs{74.72.Gh,74.40.Kb, 71.10.Hf, 74.62.-c}

\maketitle
\section{Introduction}\label{Sec:Intro}
Non-Fermi liquid (NFL) or strange metal phase is characterized by deviations from the well-defined Fermi-liquid (FL) predictions of various low-temperature properties of metals.\cite{Stewart,Sachdevreview,twodomes,PWolfle,Coleman,Taillefer,Matsuda,NFLTc} Two emergent phenomena are often observed in the NFL regime: It dissects the phase diagram between an ordered phase and the FL state, separated by a Hertz-Millis type quantum critical point (QCP);\cite{HertzMillis} secondly, superconductivity, if present, usually possesses an optimum transition temperature ($T_c$) in the NFL region. Moreover, systematic studies in various superconducting (SC) families have revealed that $T_c$ increases as the exponent $n$ decreases, i.e., as the system deviates farther from the FL behavior.\cite{Stewart,Taillefer,Matsuda,NFLTc,twodomes} It is noteworthy that considerable counter-examples are also present where a NFL phase is present without an underlying quantum phase transition\cite{Maple,twodomes,HFNFLwoQCP,CeCoIn5,Pnictide_ARPES} and/or without superconductivity, and vice versa.\cite{NFLwoSC} For these reasons, NFL state is considered an important problem towards the understanding of quantum phase transitions and unconventional superconductivity. 

The FL and NFL behavior are distinguished by multiple physical parameters. In transport, we distinguish between the FL and NFL behavior by a resistivity ($\rho$) - temperature ($T$) dependence as $\rho \rightarrow  T^2$ and $\rho \rightarrow T$, respectively. In the single particle spectrum, they are distinguished by the frequency ($\omega$) dependence of the imaginary part of the self-energy ($\Sigma^{\prime\prime}$) to be as $\omega^2$ and $\omega$ (marginal FL description), respectively. Simplified theories find a direct correspondence between the two behavior by assuming that the scattering rate ($\tau$) for resistivity solely comes from its short-lifetime as $\tau \propto 1/\Sigma^{\prime\prime}$. Applying the scaling analysis at low-temperatures $\omega\sim T$, we find that a FL transport behavior implies a long-lived, coherent quasiparticles behavior, while the NFL resistivity means incoherent many-body states. \cite{Sondhi,Sachdevbook,Vojta,PWolfle,ChubukovMaslov,ChubukovAbanov,Chubukov_singular,Sachdev_QCP,Sachdev_singular}

Such a simplified picture fails to explain several experimental features in cuprates as well as in other correlated materials. For example, it is observed that the transition from the NFL to FL state is adiabatic, i.e., at a given temperature, the resistivity exponent $n$ changes {\it continuously} from 2 to 1 or even below 1 with doping, pressure etc.\cite{Stewart,Sachdevreview,twodomes,PWolfle,Coleman,Taillefer,Matsuda,NFLTc,Gegenwart} A recent angle-resolved photoemission spectroscopy (ARPES) experiment observed a strong $k$-dependence of the self-energy in La-based cuprate.\cite{Cuprate_ARPES} It was found that the inverse of the quasiparticle lifetime ($\propto \Sigma^{\prime\prime}$) changes from $\omega^2$ to $\omega$ dependence as we move from the nodal to antinodal regions in the same sample. Moreover, the NFL self-energy persists to the overdoped sample where the transport data suggest a simple FL behavior. Again, angle-dependent magnetoresistance (ADMR) measurements on overdoped Tl-based cuprate also exhibited the similar behavior, in that the scattering rate changes from $T^2$ to $T$ behavior as we move from the nodal to the antinodal region of the sample.\cite{ADMR1,ADMR2} Recently, coexisting NFL and FL state is also observed in heavy-fermion system.\cite{YbRh2Si2}

There exist several schools of theories for the descriptions of the NFL behavior in correlated systems, which can be classified based on their assumed correlation strength. Within the Hertz-Millis theory of quantum phase transition,\cite{HertzMillis} as a system approaches a QCP, quantum fluctuations between two order parameters become {\it massless}, and the electron - (massless) boson coupling drastically suppresses the `quasiparticle' lifetime to the NFL limit.\cite{Sondhi,Sachdevbook,Vojta,PWolfle} There exists a number of perturbative approaches of the self-energy calculation based on QCP,\cite{ChubukovMaslov,ChubukovAbanov,Chubukov_singular,Sachdev_QCP,Sachdev_singular,FLEX,KontaniReview}, nearly antiferromagnetic model,\cite{NAFL,DasENFL} spin-fluctuation models,\cite{MoriyaUeda,PinesMillis} large-$N$ expansion of bosonic field,\cite{largeN} $\epsilon$-expansion of the bare dispersion,\cite{epsilonexpansion} dimension regularization\cite{dimensionregularization} method. These methods often suggest that the self-energy becomes non-analytic at the critical point, and quasiparticles can no longer be defined (in fact, in some cases, the perturbative theory itself becomes inapt at the QCP\cite{Chubukov_singular,Sachdev_singular,SSLeeReview}). On the other hand, in the strong coupling limit, one approaches the NFL limit from the other side, i.e., one basically studies how localized electrons gradually become conducting via many-body effects. A number of non-perturbative treatments, such as spin-Fermion model\cite{spinFermion}, two-fluid model,\cite{twofluid} slave-boson,\cite{slaveboson} $t-J$model,\cite{tJ}, fractional FL,\cite{OrthogonalFL,FLStar} hidden FL,\cite{HFL} DMFT\cite{DMFTQCP,Tremblay} holographic NFL,\cite{Holography} dimension-regularization method\cite{SSLee} are used here. Conductions borne out from the localized states via quantum fluctuations between the localized and conducting states. Both approaches, however, indicate a commonality that in the NFL state, the low-energy conducting states are neither fully itinerant, nor fully localized but reside in a dissonant state between them. Such a dual nature of electrons is the characteristics of the intermediate coupling region where the correlation strength is of the order of its kinetic energy term. In this correlation limit, the quantum fluctuations become either {\it massless}, or {\it marginal} and produce the imaginary part of the self-energy $\Sigma''\propto {\rm max}(|\omega|,T)$. Hence, a marginal FL (MFL) state arises in the low-$T$ limit.\cite{MFL,QMCRaja} 
\begin{figure}[t]
\hspace{-0.5cm}
\centering
\rotatebox[origin=c]{0}{\includegraphics[width=1.05\columnwidth]{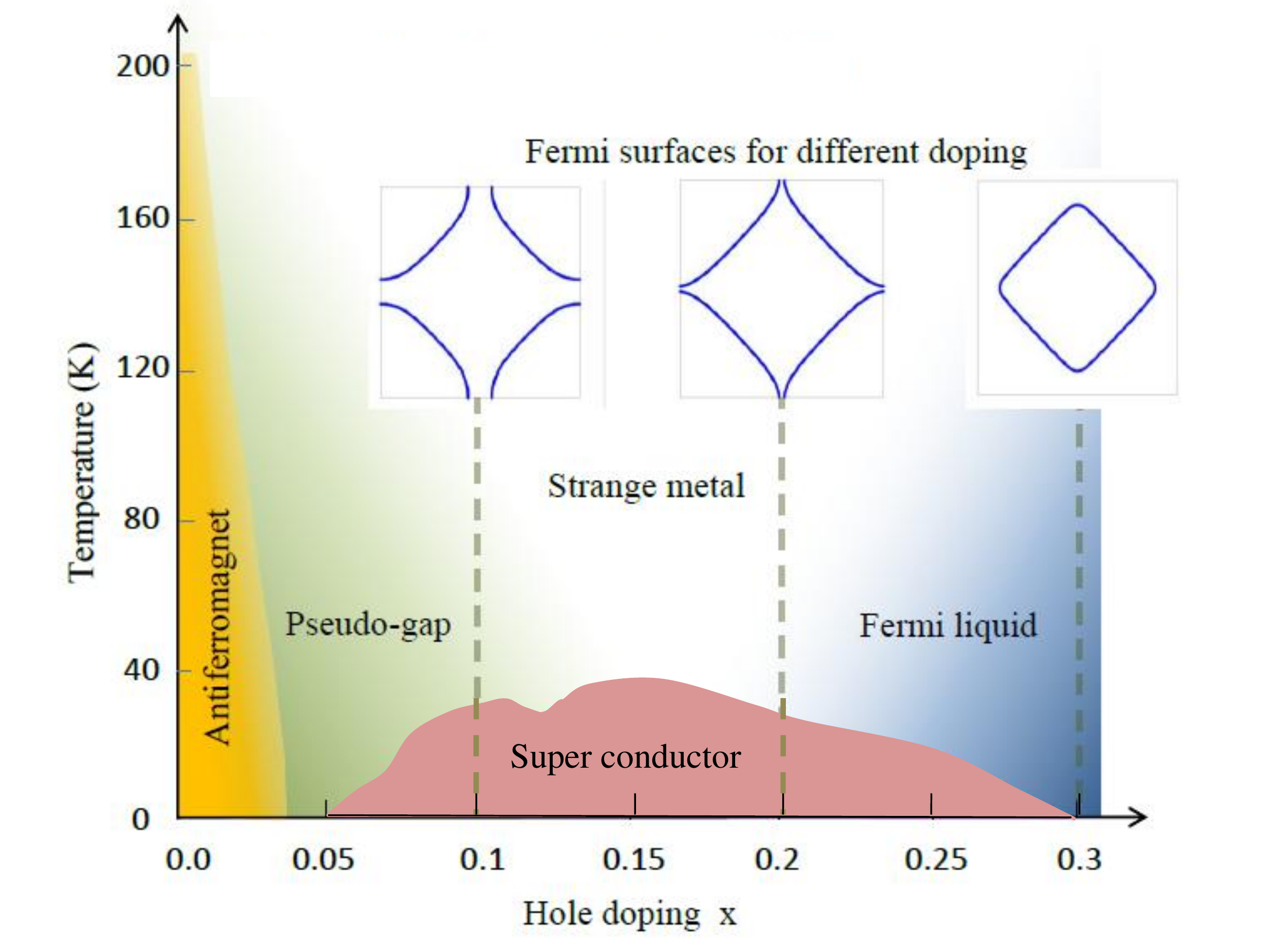}}
\caption{Schematic phase diagram of LSCO, showing the evolution of the Fermi surface across the NFL state and superconductivity. Around $x\sim 0.2$, the topological Fermi surface transition occurs where the VHS crosses the Fermi level, and the corresponding resistivity exponent becomes also minimum.
}\label{fig1}
\end{figure}

What is the correct correlation strength of cuprates? Quantum Monte-Carlo (QMC),\cite{IntCoupQMC} dynamical mean-field theory (DMFT),\cite{IntCoupMillis,IntCoupKotliar} and random-phase approximation (RPA) based fluctuation-exchange theory\cite{DasAIP,IntCoupMarkiewicz} consistently suggest that cuprates lie in the intermediate correlation strength, at least in the doped samples. The development of NFL phase in the optimal doping region, as shown in Fig.~\ref{fig1}, is studied extensively in cuprates.\cite{FLStar,PinesMillis,Kivelson,OrthogonalFL,SSLeeReview,twodomes,KontaniReview} Cluster-based calculations of QMC,\cite{CQMC} DMFT,\cite{CDMFT}  FLEX,\cite{KontaniReview,Kontani_FLEX} as well as other methods\cite{SSLee_kdep} also indicated that the self-energy is anisotropic in cuprates. In most of these methods, however, the self-energy correction arises from the antiferromagnetic (AF) fluctuations and thus dominate at the magnetic `hot-spots' where the Fermi surface (FS) crosses the magnetic BZ. Such low-energy fluctuations dominate at the magnetic QCP near 5-7\% doping, and cease to have any considerable contribution to the NFL state at the optimal doping, which is our present focus. Moreover, angle-dependence studies of resistivity,\cite{ADMR1,ADMR2} and photoemission spectroscopy\cite{Cuprate_ARPES} exhibited a different momentum dependence. It is found that both the scattering rate $\tau$, and self-energy $\Sigma''$ vary from its NFL characteristics in the antinodal region to the FL behavior in the nodal region, with no considerable change at the AF `hot-spot'. Furthermore, it is also found that the NFL like self-energy at the antinodal region survives up to the overdoped region, where the resistivity shows a global FL-behavior. In addition, it is also observed that the FS is coherent in both NFL and FL states. Therefore, the leading questions concerning the mechanism of the NFL state at the optimal doping, the analytic behavior of the self-energy in the entire $k$-space and doping, adiabatic transition with a wide region of the coexistence between the NFL and FL states in both spectroscopic and transport properties have so-far remained open.

Here we compute the {\it momentum} dependent self-energy due to density-density fluctuations within a single band Hubbard model. More specifically, the self-energy is calculated based on the momentum-resolved density-fluctuation (MRDF) model within the self-consistent RPA and fluctuation-exchange approximation.\cite{DasAIP,MRDFActinides,MRDFTMDC,MRDFNickelate,DasMottAFM} The self-energy arises due to coupling of electrons to the full spectrum of both self-energy renormalized charge and spin fluctuations within a self-consistent scheme. Charge and spin excitations have different energy and momentum scales in cuprates, and thus dominate in different doping regimes. The small-angle charge fluctuations are present near ${\bf q}\sim(\pi/4,0), (0,\pi/4)$, and are considerably weaker than the spin channels. Spin fluctuations have mainly two parts: the low-energy AF fluctuations dominating near the ${\bf Q}=(\pi,\pi)$, and (marginal) paramagnons at high energy along the ${\bf q}=(\pi,0)/(0,\pi)$ directions. In the large ordering limit ($Q>2k_F$), Sachdev {\it et al.} have shown that the paramagnons become decoupled from the AF fluctuations.\cite{Sachdev_paraAF,Sachdevbook} The AF fluctuations dies off around the AF QCP near 5\% hole doping, and do not survive up to the optimal dopings.\cite{Christossf,MarcJulien,SDWQCP}

We find here that the dominant contributions to the NFL state at the optimal doping come from self-energy dressed density fluctuations. Such density fluctuations are marginal, occur in the energy range of $300-500$meV, and survive at all dopings up to overdoped samples.\cite{paramagnonRIXS} The origin of these fluctuations is quite intriguing, and varies depending on how the self-consistency in the two-particle correlation function is treated. In our self-consistent scheme, the self-energy correction splits the electronic states into three main energy scales\cite{Markiewiczwaterfall,DasAIP,CQMC,IntCoupKotliar,IntCoupMarkiewicz,DasEPL}: there are two incoherent, localized states outside the bare band bottom and top, namely, lower and upper Hubbard bands (L/UHBs), respectively, and an itinerant band near the Fermi level containing a renormalized van-Hove singularity (VHS). These renormalized collective excitations arise form the fluctuations between the itinerant densities (concentrated mainly at the VHS), and localized densities (at the Hubbard bands). The VHS is present in cuprates near ${\bf k}\sim (\pi,0)/(0,\pi)$, while the L/UHBs are present around the $\Gamma$, and ${\bf k}\sim (\pi,\pi)$ points, respectively. Therefore, the resulting pagamagnons dominate in the region of ${\bf q}\rightarrow (\pi,0)/(0,\pi)$. Importantly, since the itinerant and localized states are always separated by the so-called `waterfall' energy ($\sim$500 meV), the fluctuations never become massless, not even at the optimal dopings. The itinerant and local density fluctuations induced self-energy thereby dominates in the antinodal region of the BZ and have its maximum effect when the VHS passes through the Fermi level (at the Lifshitz transition). Note that due to the self-energy correction and the momentum anisotropy, the VHS does not have a true singularity, rather a broad hump. Therefore, neither the density fluctuations, nor the spectral functions possess any non-analytic behavior at all dopings, and the complex self-energy remains analytical at all momenta, energy, and doping in the present model.   

We can conveniently encode the anisotropy in the self-energy by a $k$-dependent exponent ($p_{\bf k}$) in the imaginary part of the self-energy $\Sigma^{\prime\prime}$, as
%
\begin{eqnarray}
\Sigma^{\prime\prime}({\bf k},\omega)&=& \alpha_{\bf k}|\omega|^{p_{\bf k}}.
\label{Sigmak}
%
\end{eqnarray}
%
The quasiparticle residue is defined as $Z_{\bf k}=(1-\partial\Sigma^{\prime}/\partial\omega)^{-1}_{\omega=0}$, where $\Sigma^{\prime}({\bf k,\omega})$ is the real part of the self-energy. Due to analyticity of the self-energy, both the real and imaginary parts of the self-energy are related to each other by Kramers-Kronig relation at all ${\bf k}$-points. In what follows, both $p_{\bf k}$ and $Z_{\bf k}$ have characteristically similar and strong ${\bf k}$-dependence in the BZ: in the antinodal region (NFL `hot-spots') $p_{\bf k} \rightarrow 1$, and $Z_{\bf k}\rightarrow 0.3$, giving NFL self-energy, while the remaining low-density region (`cold-spots') gives $p_{\bf k}\sim 2$, $Z_{\bf k}\sim 0.8$ (see Fig.~\ref{fig3}). This allows a coexistence and competition between the NFL and FL physics in the same system. We stress that $\Sigma^{\prime\prime}({\bf k}_F,\omega=0)=0$ at all dopings, implying that all quasiparticles in the BZ (including in the NFL region) have well defined poles on the FS. However, due to the strong momentum dependence of $Z_{\bf k}$, the spectral weight gradually decreases in the antinodal region, giving the impression of a `Fermi arc' in the spectral weight maps. Such a momentum dependence of $p_k$ obtained in our MRDF method is in qualitative agreement with a QMC calculation of single band Hubbard model\cite{JarrellEPL}.

We calculate the dc resistivity using the Kubo formula. We find that the dc conductivity indicates a `global' NFL behavior, i.e., resistivity-$T$ exponent becomes $n\sim 1$, when the NFL self-energies at the antinodal region dominates over the FL self-energy in the nodal region. This occurs near the optimal doping as VHS reaches the Fermi level. Away from this characteristic doping, the phase space volume of the NFL self-energies decreases, and thus the global properties of the system gradually shift to FL-like. We note that in both NFL and FL states in the transport behavior, both NFL and FL self-energies are present on the BZ, only their relative phase space volumes change. The spectral weight transfer between the NFL to FL regions, caused by doping, temperature, and other parameters, manifests into an adiabatic transition between the FL and the NFL state in the bulk properties, such as the resistivity-temperature exponent. 

We also discuss the materials dependence of the NFL-strength and its implication to their corresponding optimum $T_c$. Pavarini {\it et al.}\cite{Pavarini} showed that the cuprates with higher $T_c$ have higher next nearest neighbor hopping element $t^{\prime}$. It is also known that as $t^{\prime}$ increases the amount of degeneracy at the VHS also increases. This, according to our calculation, gives higher strength of the NFL state, i.e. lower values of $n$. Therefore, our calculation also provides a microscopic origin to the intriguing association between the NFL state and superconductivity. We note that the results are applicable to a wider class of correlated materials in which large density of states is caused by VHS, or Liftshitz points (as in pnictides), spin-orbit coupling (in heavy-fermions and actinides) and leads to strong anisotropic self-energy effects.\cite{MRDFActinides,DasAIP}


The rest of the paper is organized as follows. In Sec.~\ref{Sec:Model}, we discuss the MRDF model and the tight-binding dispersion. Momentum-dependent self-energy result is discussed in Sec.~\ref{Sec:Selfenergy}. The overall FL/NFL behavior of a given system, characterized by the resistivity calculation and its doping dependence are discussed in Sec.~\ref{Sec:Resistivity}. In Sec.~\ref{Sec:Materials}, we study the materials dependence of the resistivity-temperature exponent and its dependence with superconducting transition temperature is presented. Finally, we discuss the advantage and limitation of our calculation in Sec.~\ref{Sec:Conclusion}, followed by conclusions. The robustness of the results against the value of the Hubbard interaction $U$ is demonstrated in the Appendix~\ref{Apx:U}.

\section{MRDF model}\label{Sec:Model}
Cuprate is a prototype of correlated superconducting family where the interplay between NFL, unconventional superconductivity, and various intertwined orders leads to a complex doping dependent phase diagram (see Fig.~\ref{fig1}).\cite{Keimer,Kivelson} Yet, the non-interacting band structure is rather straightforward with a single and strongly anisotropic band passing through the Fermi level. We consider a realistic band structure including up to fourth order tight-binding hoppings ($t$, $t'$, $t''$, and $t'''$) as $\xi_k = -2t(\cos k_x + \cos k_y)-4t^{\prime}\cos k_x\cos k_y - 2t^{\prime\prime}(\cos 2k_x+\cos 2k_y)-4t^{\prime\prime\prime}\cos 2k_x\cos 2k_y - \xi_{\rm F}$. The second nearest neighbor hopping $t^{\prime}$ has a special importance in cuprates as it controls the flatness of the band near $k=(\pi,0)$ and its equivalent points. This generates a paramount degeneracy in the DOSs, and hence VHS arises. As $t^{\prime}$ increases, the degeneracy also increases, and the system becomes more NFL like. Interestingly, an earlier Density Functional Theory (DFT) calculation demonstrated that the optimal $T_c$ in different cuprates scales almost linearly with the corresponding $t^{\prime}/t$ ratio.\cite{Pavarini} This produces a link between the NFL physics and $T_c$ with a single, {\it ab-initio} parameter. 

\begin{figure}[t]\label{Feynmann_diagram}
\centering
\rotatebox[origin=c]{0}{\includegraphics[width=0.99\columnwidth]{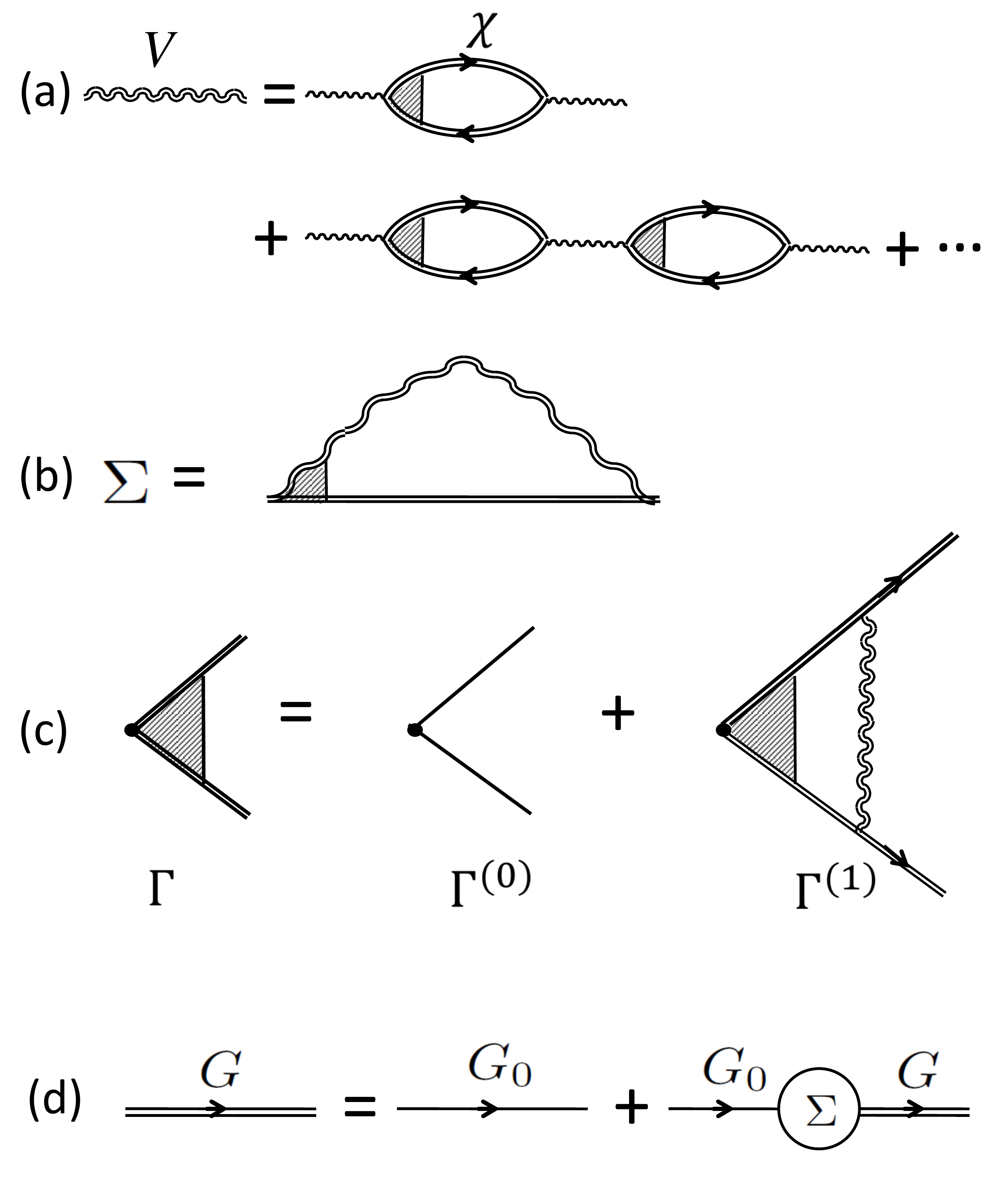}}
\caption{Diagrams of various quantities of the MRDF model. (a) MRDF potential, (b) self energy, (c) Bethe-Salpeter vertex equation, (d) Dyson equation. Double solid line represents self-energy dressed Green's function $G$, while single solid line is for the bare Green's function ($G^0$). Double wavy line represents the fluctuation-exchange potential. $\Gamma^{(0)}$, and $\Gamma^{(1)}$ are the bare and self-consistent vertex corrections (the same diagram applies to both density and current vertex corrections). }
\label{fig2a}
\end{figure}

Our starting point is a single band Hubbard model. The present MRDF model is restricted to the intermediate coupling model, where the value of Hubbard $U$ is just below the self-energy renormalized bandwidth $\mathcal{W}$ (evaluated self-consistently). This is the Brinkman-Rice criterion.\cite{BR} The value of $U$ determines the overall strength of the NFL state, but interestingly, it does not affect much the anisotropy in the self-energy (as shown in Appendix~\ref{Apx:U}). We only take into account the correlation part of the Hubbard model, and compute the full spectrum of both charge- and spin-fluctuations in a self-consistent way. The correlation part is included within the RPA approximation, by summing over the bubble diagrams (see Fig.~\ref{fig2a}), where the ladder diagrams are included in the Bethe-Salpeter vertex correction.\cite{BetheSalpeter} The higher-order Maki-Thompson (MT),\cite{MT} and Aslamasov-Larkin (AL)\cite{AL} terms, beyond the RPA diagram, are shown in Appendix~\ref{Apx:Cond} to scale as $U/\mathcal{W}^2$, and $U^2/\mathcal{W}^6$, respectively, and thus can be neglected in the intermediate coupling regime. The coupling between density fluctuations and electrons gives rise to a complex self-energy, which can be calculated within the Hedin's approach.\cite{Hedin} Here we use the self-consistent momentum-resolved density fluctuation (MRDF) method\cite{DasAIP,MRDFActinides,MRDFTMDC,MRDFNickelate} in which all quantities including single-particle Green's function, two-particle correlation functions, and the three-point vertex corrections are calculated self-consistently with the self-energy correction. In this way, the present method is an improved version of the FLEX model\cite{FLEX} without self-energy corrections in the two-particle term, or the single-shot GW calculation without a vertex correction.\cite{GW,GWwoVertex} The Hedin's self-energy in terms of the self-energy dressed spectral function $A$ can be written as (see appendix~\ref{Apx:Calculation}): 
\begin{eqnarray}
&&\Sigma_{\nu}({\bf k},\omega)=\frac{1}{N}\sum_{\bf q}\int_0^{\infty}\frac{d\varepsilon}{2\pi}\int_{-\infty}^{\infty}\frac{d\omega^{\prime}}{2\pi}V_{\nu}({\bf q},\varepsilon)\Gamma_{\nu}({\bf k},{\bf q},\omega',\epsilon)
\nonumber\\
&&\times A({\bf k}-{\bf q},\omega^{\prime})\left[\frac{1-f({\omega^{\prime}})+n(\varepsilon)}{\omega+i\delta-\omega^{\prime}-\varepsilon}+\frac{f({\omega^{\prime}})+n(\varepsilon)}{\omega+i\delta-\omega^{\prime}+\varepsilon}\right],
\label{selfenergy}
\end{eqnarray}
where $f(\omega)$ and $n(\varepsilon)$ are fermionic and bosonic distribution functions, respectively. $N$ is the total number of lattice sites. $A({\bf k},\omega)=-{\rm Im}G({\bf k},\omega)/\pi$ and $G({\bf k},\omega)=\left[\omega-\xi_{\bf k}-\Sigma({\bf k},\omega)\right]^{-1}$ are the self-energy dressed spectral weight and Green's function, respectively.  $V_{\nu}({\bf q},\varepsilon)$ is the back-reaction potential of quasiparticle density fluctuations which are separated into the spin ($\nu=1$) and charge ($\nu=2$) density channels within the RPA model as
\begin{eqnarray}\label{V_pot}
V_{\nu}({\bf q},\varepsilon) = \frac{\eta_{\nu}}{2}{\rm Im}\left[\frac{U^2 \chi({\bf q},\varepsilon)}{1\mp U \chi({\bf q},\varepsilon)}\right],
\label{Vq}
\end{eqnarray}
where $\eta_1=3$, and $\eta_2=1$, and $U$ is the onsite Hubbard interaction. $\chi$ is the corresponding bare correlator, evaluated self-consistently, as 
\begin{eqnarray}
\chi({\bf q},\varepsilon)&=&\frac{1}{N}\sum_{\bf k}\int\frac{d\omega_1}{2\pi}\int\frac{d\omega_2}{2\pi}A({\bf k},\omega_1)A({\bf k}+{\bf q},\omega_2)\nonumber\\
&&\qquad\times \Gamma({\bf k},{\bf q},\omega_1,\omega_2)\frac{f(\omega_1)-f(\omega_2)}{\varepsilon+i\delta-\omega_2+\omega_1}. 
\label{chi}
\end{eqnarray}
Here $\Gamma({\bf k},{\bf q},\omega_1,\omega_2)$ is the density vertex correction. We note that due to the strong anisotropy in the self-energy, the Midgal's approximation is not valid here, and vertex correction becomes important.
\begin{figure}[t]\label{fluctuation_spectrum}
\centering
\rotatebox[origin=c]{0}{\includegraphics[width=0.99\columnwidth]{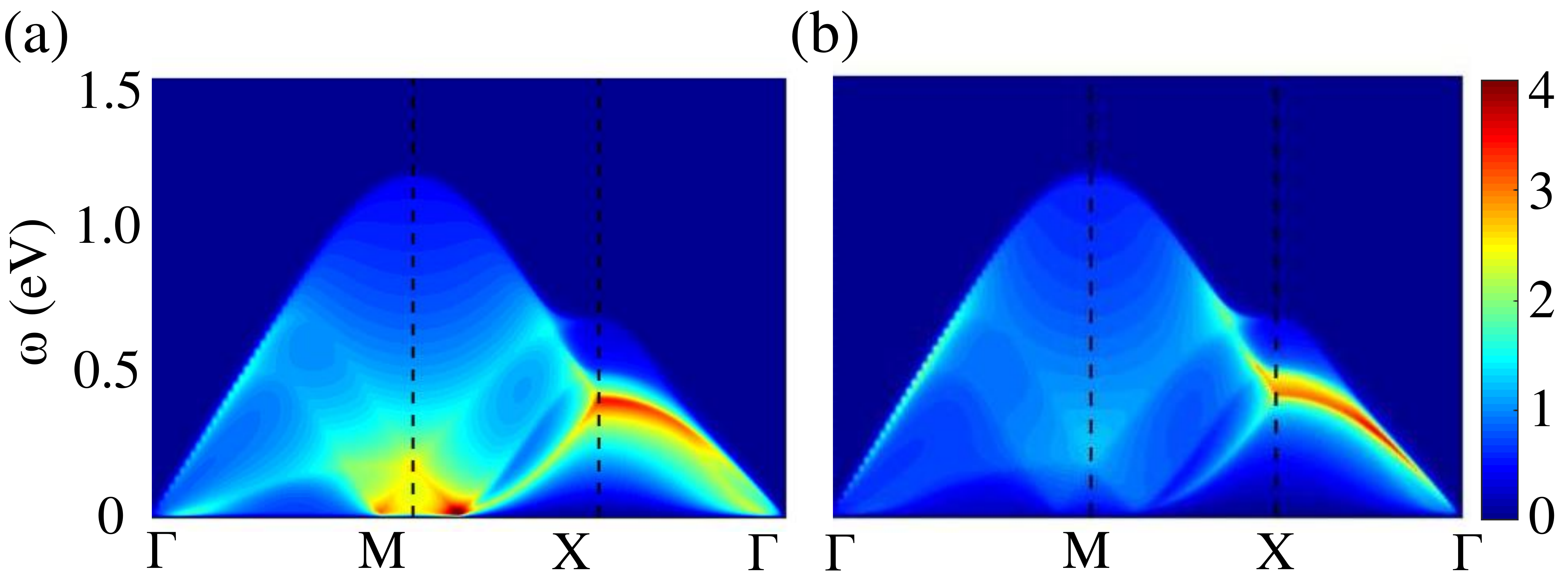}}
\caption{Density fluctuation spectrum for (a) spin, (b) charge channels. Here we plotted the imaginary part of the RPA susceptibilities: $\chi/(1\mp U \chi)$ for spin and charge densities  as a function of $\omega$ along three high symmetric momentum directions: $\Gamma$(0,0)-M$(\pi,\pi)$-X$(\pi,0)$-$\Gamma$ for 20\% hole doped LSCO. These RPA susceptibilities are directly linked to the fluctuation potential $V_{\nu}$ in Eq.~\eqref{V_pot} with the multiplication of the constant factor $U^2\eta_{\nu}/2$. The main feature of the density fluctuation is the dispersive paramagnon-like mode along the X - $\Gamma$ direction. Such a mode is observed in RIXS spectrum in various cuprates.\cite{RIXS} This mode is responsible for the NFL state in the antinodal region.}
\label{densityfluc}
\end{figure}

Again, the ${\bf k}$-dependent $\Sigma({\bf k},\omega)$ prioritizes the current-current vertex term ${\bf \Gamma}$, which also affects the density vertex $\Gamma$ due to conservation principles (it is customary to denote the current and density vertices by vector and scalar symbols ${\bf \Gamma}$, and $\Gamma$, respectively)\cite{CVC}. Since the system possesses both gauge- and spin-rotational symmetries without and with the self-energy corrections, the conservations of charge and spin densities lead to a simplified algebraic form of the vertex correction, as known by Ward's identity.\cite{Ward} This identity imposes a specific relation between the self-energy and the density vertex correction as (see Appendix ~\ref{Apx:Vertex} for the derivation) \cite{def_sus}
\begin{eqnarray}
\Gamma({\bf k},{\bf q},\omega,\epsilon) 
\approx 1-\frac{\partial \Sigma'({\bf k},\omega)}{\partial \omega}=Z^{-1}_{{\bf k}}(\omega).
\label{vertex}
\end{eqnarray}
Such a vertex correction is not only important to preserve the sum-rules, but also it helps to produce the correct frequency values ($\sim 500$meV) and the strength of the correlation functions, $V$, the self-energy $\Sigma$, as well as spectral functions $A$, in consistence with their corresponding experimental results.\cite{paramagnonRIXS}

While the numerical computations involve the full self-energy anisotropy, some interesting properties can be extracted if we impose the FL ansatz of the self-energy. That means, we approximate the self-energy as $\Sigma({\bf k},\omega)=\Sigma({\bf k},0)+ (1-Z_{\bf k}^{-1})\omega$, where $Z_{\bf k}$ is the anisotropic quasiparticle residue at the Fermi level, defined before. We obtain the dressed quasiparticle band as $\bar{\xi}_{\bf k}=Z_{\bf k}\xi_{\bf k}$.  Substituting the corresponding spectral function as $A({\bf k},\omega)=Z_{\bf k}/(\omega+i\delta-\bar{\xi}_{\bf k})$ in Eq.~\eqref{chi}, we find that $\chi({\bf q},\varepsilon)=\Gamma Z^2\chi_0({\bf q},\varepsilon)=Z\chi_0({\bf q},\varepsilon)$, where $\chi_0$ is the bare Lindhard susceptibility (without a self-energy correction), and $Z$ is the momentum averaged renormalization factor. This means, both the kinetic energy and the correlation function are renormalized in the same way, a consequence of the the Ward's identity. Furthermore, the MRDF potential in Eq.~\eqref{Vq} is also renormalized by the same value if the interaction $U$ is also renormalized similarly, i.e., if $U=ZU_0$, where $U_0$ is the bare Hubbard $U_0$. This yields $V_{\nu}({\bf q},\varepsilon)=ZV^0_{\nu}({\bf q},\varepsilon)$, where $V^0_{\nu}({\bf q},\varepsilon)$ is the bare fluctuation-exchange potential consisting of bare $\chi_0$, and bare $U_0$ in Eq.~\eqref{Vq}. Since the kinetic and interaction terms scale in the same way, the system always maintains the intermediate coupling strength. Once we turn on the momentum dependence of the renormalization factor, such a simple, analytical proof is difficult to achieve. However, the $f$-sum rules remained valid as shown in Sec.~\ref{Sec:Discussion}, and the MRDF method maintains the intermediate coupling scenario.

\section{Self-energy results}\label{Sec:Selfenergy}

\begin{figure}[ht]
\centering
\rotatebox[origin=c]{0}{\includegraphics[width=0.99\columnwidth]{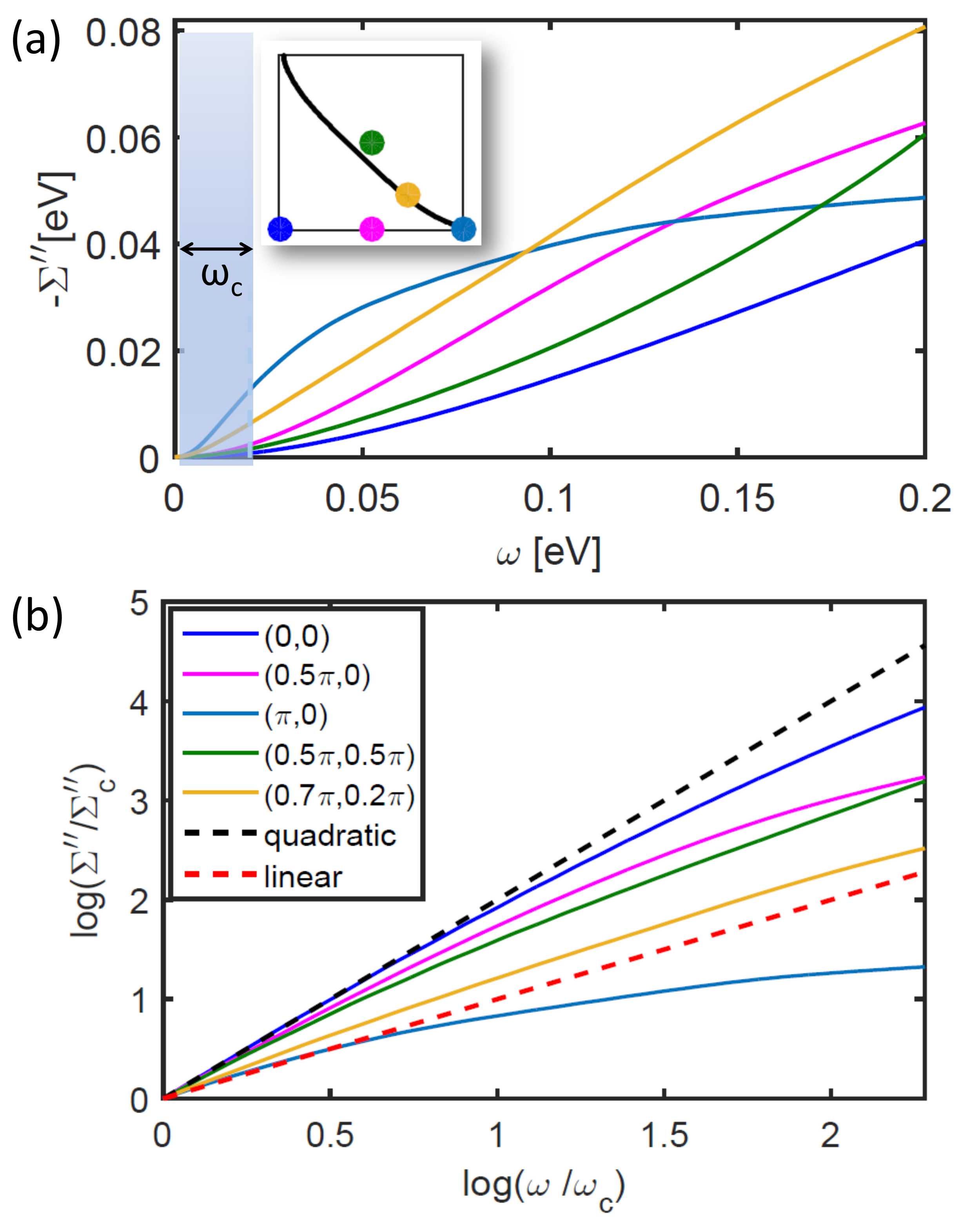}}
\caption{{(a) Calculated $\Sigma''(\omega)$ at different ${\bf k}$-points for 20 \% hole dopped LSCO. We excluded the very low frequency range of the order of impurity broadening 2$\delta$ (shaded region) to extract the exponent $p_k$ (see appendix C). {\it Inset}: Corresponding $k$-points in the first quadrant of the BZ where the self-energy is plotted. Bottom-left corner is at the $\Gamma$-point, while the top-right corner is the $(\pi,\pi)$-point. The black line indicates the non-interacting Fermi Surface. (b) Logarithmic plot of $\Sigma''(\omega)$ vs $\omega$, rescaled with $\Sigma^{\prime \prime}_{c} = \Sigma^{\prime \prime}(\omega_{c})$ and $\omega_{c}=2\delta$, respectively. Colors represent the same ${\bf k}$-points as in (a). Black and red dashed lines are guide to the eyes to a quadratic and linear behavior, respectively.
}
}\label{fig2}
\end{figure}

For the presentation of the self-energy results in this section, we focus on La$_{2-x}$Sr$_{x}$CuO$_{4}$ (LSCO) cuprate. Its tight-binding (TB) band parameters are obtained from the corresponding DFT band structure (see Table~\ref{table} below). The self-energy is shown near the optimal doping ($x=0.20$) with $U=1 eV$ (where the bandwidth is $\mathcal{W}\sim 4 eV$). The self-energy is plotted for several representative momenta in Fig.~\ref{fig2}. The results can be compared with the corresponding results obtained from ARPES for the same sample. Both experiment and theory consistently exhibit a characteristic momentum-dependence of the self-energy. $\Sigma^{\prime\prime}$ varies linearly with frequency in the antinodal region, while it gradually becomes quadratic as we move towards the nodal region. 

\begin{figure}[t]
\centering
\rotatebox[origin=c]{0}{\includegraphics[width=0.99\columnwidth]{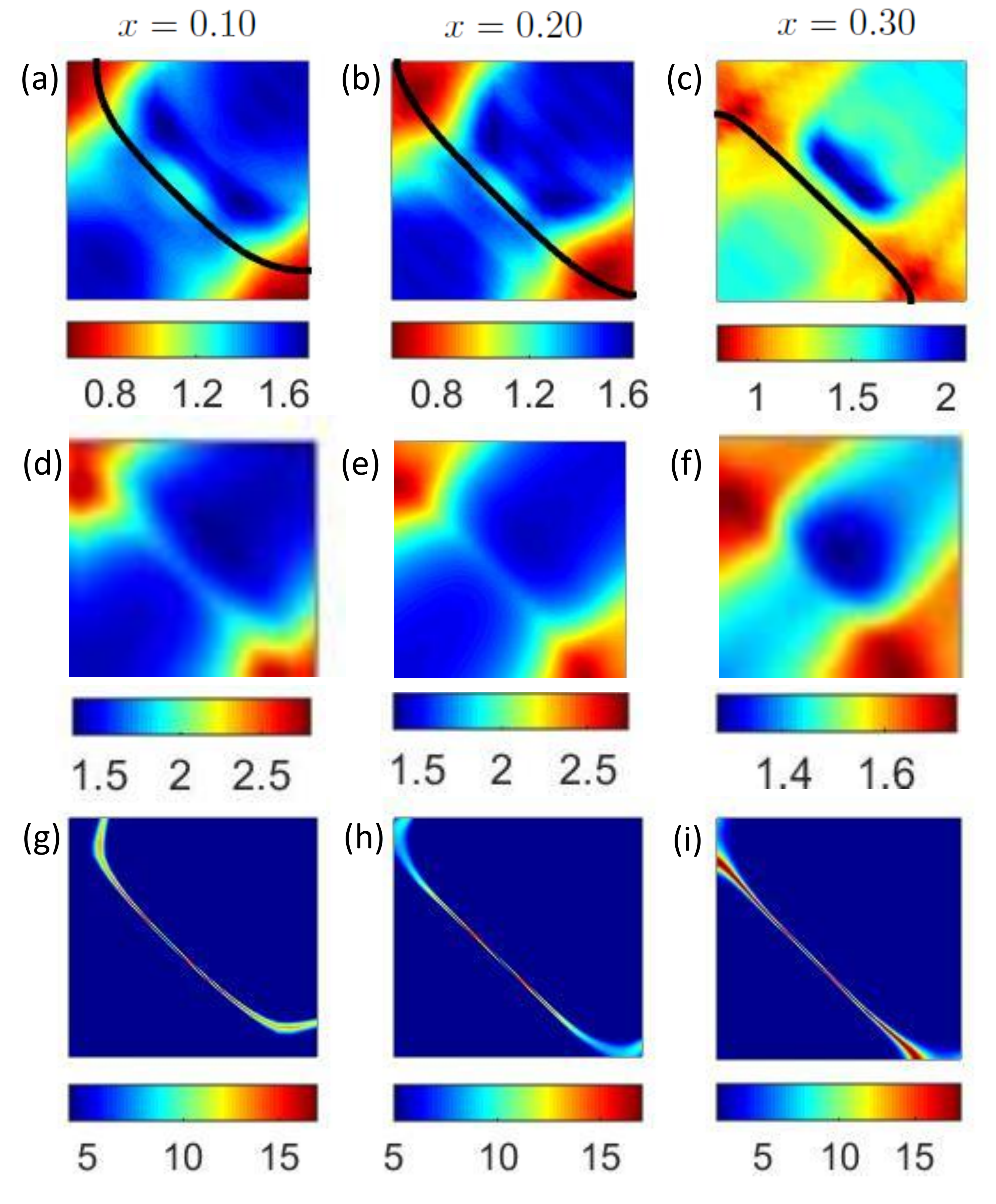}}
\caption{{(a-c) We show the self-energy exponent $p_k$ in the entire first quadrant of the BZ in the underdoped, optimal doped, and the overdoped regions, respectively. Bottom-left corner is at the $\Gamma$-point, while the top-right corner is the $(\pi,\pi)$-point. The black lines indicate the corresponding Fermi Surface and the colorbars indicate the value of exponent $p_k$. (d-f) Inverse of quasiparticle weight $Z_{k}^{-1}=m^{*}/m_b$ at Fermi energy ($\omega=0$) is plotted for the same dopings in in (a). (g-i) Spectral weight plots at the Fermi level, including the momentum dependence self-energy. Each column corresponds to the same doping.}
}\label{fig3}
\end{figure}

The origin of the momentum dependence of the self-energy can be traced back to the momentum dependence of $V_{\nu}$ [Fig.~\ref{densityfluc}] and the spectral weight maps [Fig.~\ref{fig3}]. We focus the discussion on the two momentum regions: the NFL region around ${\bf k}_{\rm v}\sim (\pi,0)$, and the FL regions ${\bf k}_{\rm h}\sim \Gamma$, and $(\pi,\pi)$. The self-energy creates incoherent, localized states at the bottom and top of the bands at the $\Gamma$, and $(\pi,\pi)$ point, which are reminiscences of the lower and upper Hubbard bands (L/UHBs), respectively. The low-energy VHS states around ${\bf k}_{\rm v}$ near the Fermi level remain `itinerant'\cite{DasMottAFM,DasAIP}. These two states are separated by the so-called `waterfall' energy ($\sim$500 meV) where the spectral weight is strongly suppressed. $V_{\nu}({\bf q},\epsilon)$ arises mainly from density fluctuations between the itinerant (at VHS) and localized (at the L/UHB) states in the particle-hole channel. Below the NFL-doping where the VHS lies below $E_F$, the density fluctuations arise between the VHS at ${\bf k}_{\bf v}$ and the UHB at $(\pi,\pi)$. Above the NFL-doping, the VHS crosses above the Fermi level, and the corresponding fluctuation switches channels between the VHS and the LHB at the $\Gamma$-point. In both cases, the momentum conservation principle localizes $V_{\nu}$ at $({\bf q}_{\rm v},\epsilon_{\rm sf})$, where $\epsilon_{\rm sf}\sim 300-500$meV, and ${\bf q}_{\rm v}\sim (\pi,0)/(0,\pi)$. We have visualized the self-energy dressed density fluctuation spectrum in Fig.~\ref{densityfluc} for the spin and charge channels. Consequently, these fluctuations persists from underdoping to overdoping, as  observed by resonant-inelastic X-ray scattering spectroscopy (RIXS)\cite{RIXS}. A direct comparison of the computed density fluctuations spectrum with the corresponding RIXS data for different dopings have been shown elsewhere.\cite{DasLEK,DasAIP} Substituting $V_{\nu}({\bf q}_{\bf v},\epsilon_{\rm sf})$ in Eq.~\eqref{ImSigma}, we find that $\Sigma_{\nu}^{\prime\prime}({\bf k},\omega)\approx  V_{\nu}({\bf q}_{\rm v},\epsilon_{\rm sf})A({\bf k}-{\bf q}_{\rm v},\omega+\epsilon_{\rm sf})$. Therefore, we can relate the NFL self-energy at $\Sigma_{\nu}^{\prime\prime}({\bf k}_{\rm v},\omega)$ to depend mainly on the Hubbard states $A({\bf k}_{\rm h},\epsilon_{\rm sf}+\omega)$. In other words, the NFL self-energy arises from the `high-energy' localized Hubbard bands, which transfer the localized spectral density via density-density fluctuation channels to the low-energy states at the antinodal region. On the other hand, the FL self-energies near ${\bf k}_{\rm h}$-points depend mainly on the itinerant VHS spectral weights at $A({\bf k}_{\rm v})$. Since the spectral function has isolated poles at all moment and frequency, both the NFL and FL self-energies are analytic functions in the present case. This way the present model is different from the prior perturbative treatments of the NFL state.\cite{ChubukovMaslov,ChubukovAbanov,Chubukov_singular,Sachdev_QCP,Sachdev_singular,SSLeeReview,cuspVHS}

Exact extraction of the frequency exponent $p_k$ is hampered by the impurity broadening term $\delta$. $\Sigma^{\prime\prime}({\bf k},0)\sim 0$, so to deal with the Green function's poles, one needs to add an impurity broadening $\delta$ in the calculation. This effectively gets added to the self-energy, and changes the frequency dependences for $\omega\le 2\delta$. So, we fit $\Sigma^{\prime\prime}$ above $\omega>2\delta$ (as highlighted in Fig.~\ref{fig2}(a)), and the corresponding log-log plot is shown in Fig.~\ref{fig2}(b) (a detailed procedure is given in Appendix~\ref{Apx:SEextraction}). From the log-log plot, we can conclude that the exponent is $\sim 1$ in the antinodal region (NFL-state), and $\sim 2$ away from the antinodal region (FL-states). In addition, the fitting is not monotonic with frequency, because both the exponent $p_{\bf k}$ and the coefficient $\alpha_{{\bf k}}$ in Eq.~\eqref{Sigmak} are also frequency dependent. But for the low-temperature transport properties, the low-energy fitting suffices a good explanation.  

In Fig.~\ref{fig3} we show the momentum dependence of the exponent $p_k$, and compare it with that of the mass renormalization $m^*/m_b=Z^{-1}_{\bf k}$ ($m_b$= bare band mass), and the spectral weight map $A({\bf k}_F,0)$. The results are compared for three different dopings: at $x=0.1$ (left), optimal doping $x=0.2$ (middle), and $x=0.3$ (right). We immediately observe a one-to-one correspondence between the three quantities at all dopings, further justifying that the self-energy is always non-singular. The spectral weight can be defined in terms of $Z_{\bf k}$ and $\Sigma^{\prime\prime}$ as 
\begin{eqnarray}
A({\bf k},\omega) = -\frac{1}{\pi}\frac{Z_{\bf k}(\Sigma^{\prime\prime}({\bf k},\omega)+\delta)}{(\omega-\bar{\xi}_{\bf k})^2 + (\Sigma^{\prime\prime}({\bf k},\omega)+\delta)^2}.
\end{eqnarray}
Since $\Sigma^{\prime\prime}=0$ at $\omega=0$ at all ${\bf k}$, we can approximate the spectral functions as $A({\bf k},\omega\rightarrow 0)=Z_{\bf k}\delta(\omega-\bar{\xi}_{\bf k})$, where $\bar{\xi}_{\bf k}=Z_{\bf k}[\xi_{\bf k}+\Sigma^{\prime}({\bf k},0)]$. This suggests that the FS remains coherent at all momenta and dopings. The self-energy dressed FS deviates from the bare FS (black line) both in shape and spectral weight. The spectral weight renormalization on the FS is solely governed by the quasiparticle residue $Z_{\rm k}$. The shift of the FS is dictated by $\Sigma^{\prime}({\bf k},0)$ which is also related to $p_k$ via Kramer's-Kronig relation:
\begin{equation} 
\Sigma^{\prime}({\bf k},0) = \frac{1}{\pi}\int_{-\infty}^{\infty} d\omega \frac{\Sigma^{\prime\prime}({\bf k},\omega)}{\omega}.
\label{Sigmak0}
\end{equation} 
Therefore, we observe that the renormalized Fermi momenta ${\bf k}_F$ deviate more from its non-interacting values in the antinodal direction, compared to the other points. Finally, the number of electron is kept fixed by recalculating the chemical potential $\xi_{F}$ with the self-energy correction. Therefore, the Luttinger theorem remains valid at all dopings. 

The above analysis demonstrates that due to the analytic form of the self-energy, $p_{\bf k}$, $Z_{\bf k}$, and $A_{\bf k}$ all are related to each other at all ${\bf k}$-values. All three are minimum at the antinodal point, suggesting that the states near this region are more strongly correlated than the rest of the BZ. Also, from Eq.~\eqref{Sigmak0}, we find that $\Sigma^{\prime}({\bf k},0)$ is maximum at the antinodal point, and thus the corresponding Fermi momenta ${\bf k}_F$ deviate more from its non-interacting values here. To have the Luttinger theorem valid, the Fermi momenta elsewhere must be smaller.

\begin{figure}[t]
\centering
\rotatebox[origin=c]{0}{\includegraphics[width=0.99\columnwidth]{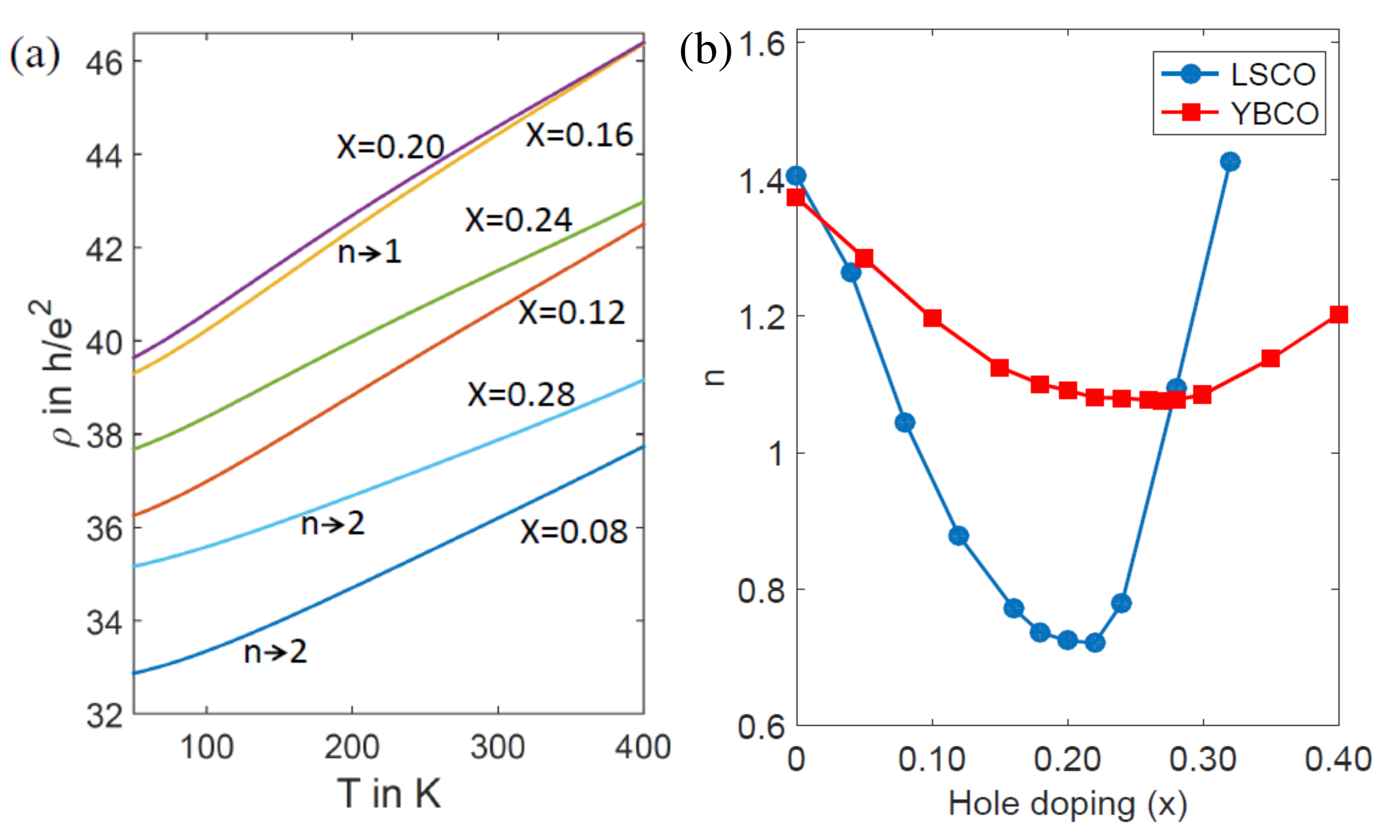}}
\caption{Resistivity is plotted as a function of temperature for different doping for LSCO. (Due to finite broadening of  the Green's function at very low frequency, we cannot determine the exponent at very low-temperature.) The corresponding exponent of each curve is indicated in the adjacent label. (b) The resistivity-temperature exponent in the low-$T$ region is plotted as a function of dopings for two different materials. Both materials exhibit minimum in exponent near the optimal dopings where the corresponding VHS passes through the Fermi level.
}\label{fig4}
\end{figure}

The overall $k$-dependence of $p_k$ remains similar at all dopings: $p_k$ attains its minimum value around the antinodal region. In the underdoped region, where the VHS is well below $\xi_{\rm F}$, we find that the overall $p_k$ profile is less $k$-sensitive. Near the optimal doping, where the VHS exactly crosses above $\xi_{\rm F}$, we find that the $k$-dependence of $p_k$ becomes strongest, and the NFL region occupies larger BZ volume. Also at optimal doping, $p_k$ obtains its minimum value of $\sim$0.65 near the antinodal region, which is the minimum possible value of $p_k$ at all dopings and momenta for this material. At this doping, we find below that the resistivity-temperature exponent also attains its minimum value of $\sim$0.7 as shown in Fig.~\ref{fig4}(b). Finally, as the VHS crosses above $\varepsilon_{\rm F}$, again the value of $p_k$ increases. Interestingly, in the overdoped region, where the resistivity data below shows an overall FL-behavior, the antinodal regions continue to show NFL self-energy behavior, in consistent with the ARPES data on LSCO at $x=0.23$.\cite{Cuprate_ARPES}

The result suggests that the quasiparticles have well-defined poles in both FL and NFL states at all ${\bf k}_F$, but owing to the ${\bf k}$-dependent $\Sigma^{\prime}({\bf k},0)$, the deviation of the poles from its non-interacting FS is not monotonic on the FS. The only source of the spectral weight renormalization on the FS is the momentum dependent $Z_k$. Expectedly, spectral weight gradually decreases as we move to the antinodal directions, giving the shape of a coherent `Fermi arc', often observed in underdoped cuprates.\cite{Fermiarc} 

\section{Resistivity calculations}\label{Sec:Resistivity}

When the $k$-dependence of the self-energy is neglected, a direct link between the microscopic single-particle spectral properties and the macroscopic transport behavior ($n\approx p$) can be established. However, as the system acquires strong anisotropy in $p_k$, it becomes less intuitive to deduce the overall correlation landscape from transport properties. We compute the DC conductivity by using the Kubo formula. We consider a one-loop (bubble diagram) with the current-current vertex correction ${\bf \Gamma}$. Because of the vertex correction, the higher-order MT,\cite{MT} and AL terms\cite{AL} for the current-current correlation functions give vanishingly small contributions, unless one enters into non-analytic self-energy\cite{Hartnollop} or if the self-energy has pseudogap behavior.\cite{Tremblayop} Such an one-loop Kubo formula, with and without vertex correction, is also used previously in cuprates within DMFT calculation.\cite{IntCoupMillis,IntCoupKotliar,HFL} The current vertex is calculated from the same Bethe-Salpeter form,\cite{BetheSalpeter} which is calculated self-consistently using Ward identity\cite{Ward} (see Appendix~\ref{Apx:Vertex}). Within the linear response theory, in the limit of ${\bf q}\rightarrow 0$, we obtain:
\begin{eqnarray}
\sigma&=&\frac{e^{2}}{3\hbar^{2}m^2}\frac{1}{N}\sum_{\bf k} {\bm \Gamma}^{(0)}({\bf k},\omega)\cdot{\bm \Gamma}({\bf k},\omega) \nonumber\\
&&\qquad\times \int\frac{d\omega}{2\pi}A^{2}({\bf k},\omega)\left(-\frac{df(\omega)}{d\omega}\right),
\label{conductivity}
\end{eqnarray}
where $e$ and $\hbar$ have the usual meanings, and ${\bm \Gamma}^{(0)}({\bf k},\omega)$ and ${\bm \Gamma}({\bf k},\omega)$ are the bare and full current vertices. For ${\bf q}\rightarrow 0$, the bare vertex reduces to ${\bm \Gamma}^{(0)}({\bf k},\omega)=m{\bf v}({\bf k})$, and the full vertex is 
\begin{eqnarray}
{\bm \Gamma}({\bf k},\omega)=m{\bf v}({\bf k}) + m{\bm \nabla}\Sigma({\bf k},\omega)= -m{\bm \nabla}G^{-1}({{\bf k}},\omega).
\end{eqnarray}
The conductivity obeys the $f$-sum rule as shown in Sec.~\ref{Sec:Discussion}. We consider $\sigma_{xx}$ components only. In the absence of any anomalous term, the resistivity is obtained as $\rho_{xx} =1/\sigma_{xx}$. 

The results are presented in Fig.~\ref{fig4}(a) for LSCO at different dopings. We find that the resistivity exponent becomes minimum near the optimal doping where the VHS crosses $\xi_{\rm F}$, see Fig.~\ref{fig4}(b). Here, the system acquires dominant NFL-behavior with $n\sim $1. At the same doping, the self-energy exponent $p_k$ in Fig.~\ref{fig3}(b) not only obtains its minimum value (min$[p_k]\sim 0.65$), but also it occupies larger $k$-space area. However, the other parts of the BZ remain FL-like with $p_k$ as large as $\sim$1.6. Similarly, in both under- and overdopings, where $n\rightarrow 2$, the antinodal region continues to have $p_k\sim 1$. Finally, we repeat the calculation for the YBCO material as a function of doping, using the corresponding realistic tight-binding parameter set,\cite{DasAIP} and the results are shown in Fig.~\ref{fig4}(b). We consistently find that $n$ is minimum near its optimal doping as the corresponding VHS passes through $\xi_{\rm F}$. Cautionary remarks are in order. We have extended the one-band model to the deep underdoped region without including the pseudogap and other competing orders. Therefore, our calculation does not represent the experimental results in the deep underdoped region.

ADMR technique has the ability to probe the angular variation of the resistivity by tilting the magnetic field with the sample orientation. This allows to effectively measure the scattering life-time $1/\tau\propto \rho$ as a function of Fermi surface angle $\theta=\tan^{-1}(k_y/k_x)$. An earlier ADMR study on overdoped Tl$_2$Ba$_2$CuO$_{6+x}$ found that $1/\tau$ varies as $T^2$ in the nodal region ($\theta=0$) and it gradually changes to $T$ in the antinodal region ($\theta=45^{o}$).\cite{ADMR1,ADMR2} This result is consistent with our findings of quasiparticle life-time variation shown in Fig.~\ref{fig3}(a-c). Note that even the overall resistivity exponent is close to 2 in the overdoped region, however, its local variation reveals that both the single-particle life-time and scattering rate consistently remain NFL-like in the antinodal region.

\section{Materials dependence of $n$ and its correlation with $T_c$}\label{Sec:Materials}

\begin{figure}[ht]
\centering
\rotatebox[origin=c]{0}{\includegraphics[width=0.99\columnwidth]{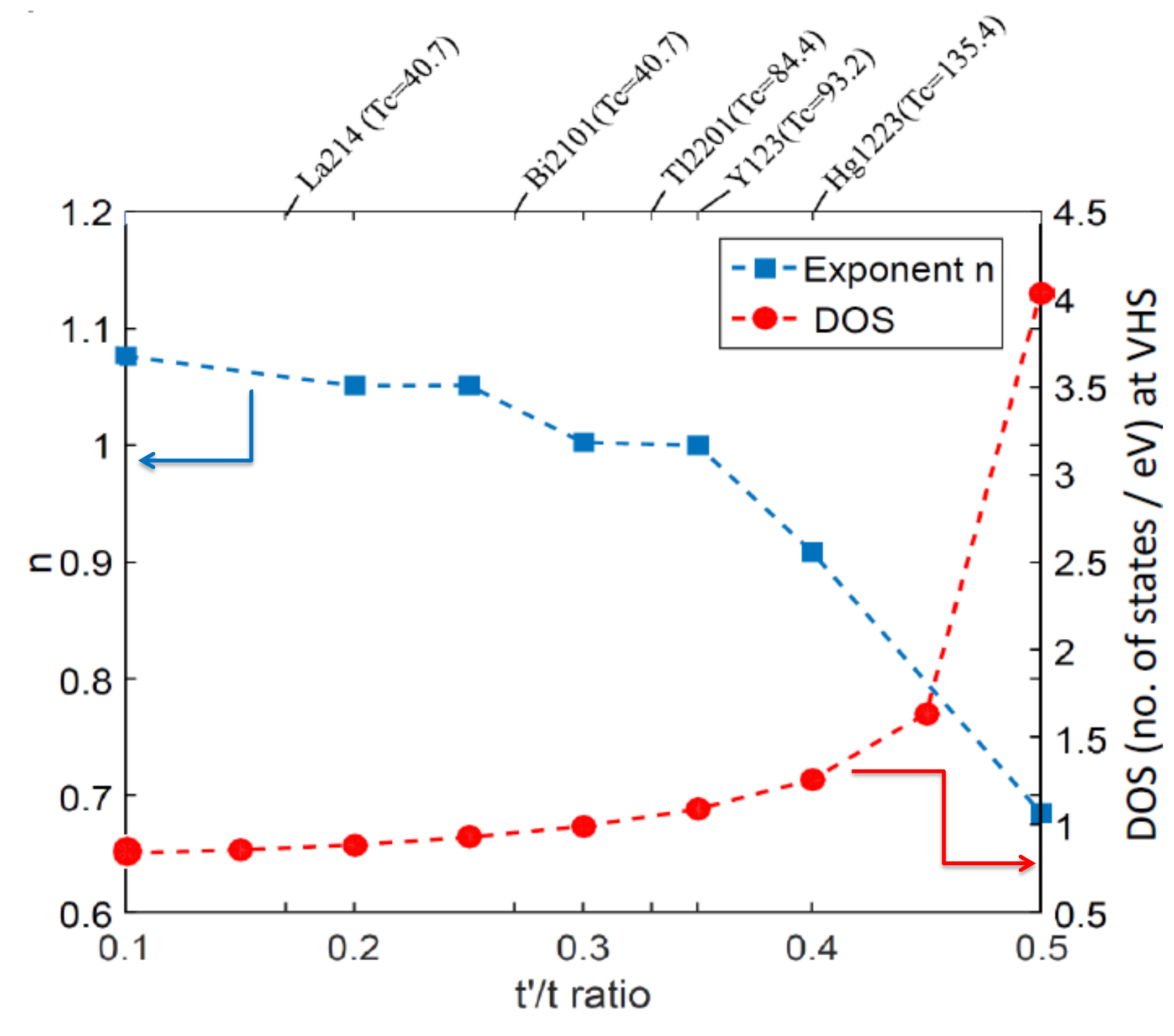}}
\caption{The resistivity exponent $n$, at the doping where the VHS passes through the Fermi level, is plotted for different values of $t^{\prime}/t$, representing different cuprate materials.\cite{Pavarini} This is the minimum value of $n$ obtained across the doping range for a given $t^{\prime}/t$ since the DOS at $E_{F}$ is maximum here (see the circle symbols for the DOS in the right-hand panel). At the top of the figure, we mention the corresponding cuprate materials with corresponding $T_{c}$, having different values of $t^{\prime}/t$ as obtained from the DFT calculation in Ref. ~\onlinecite{Pavarini}. Chemical compositions of cuprate materials are: La$_{2}$CuO$_{4}$ (La214), Bi$_{2}$Sr$_{2}$CuO$_{6}$ (Bi2101), Tl$_{2}$Ba$_{2}$CuO$_{6}$ (Tl2201), YBa$_{2}$Cu$_{3}$O$_{7}$ (Y123), HgBa$_{2}$Ca$_{2}$Cu$_{3}$O$_{8}$ (Hg1223).
}\label{fig5}
\end{figure}

The celebrated paper by Pavarini {\it et al.}\cite{Pavarini} pointed out an intriguing relationship between the $t'/t$ ratio obtained in different materials with their $T_c$. $t'/t$ triggers higher degeneracy in the DOS (see appendix B3), and hence it is natural to expect that the strength of the NFL state would also increase. We calculate the resistivity exponent $n$ for different values of $t^{\prime}/t$ by fixing the VHS at the $\xi_F$, and the result is plotted in Fig.~\ref{fig5}. Indeed, we find that with increasing $t^{\prime}/t$, $n$ decreases, that means, the system becomes more NFL like. With increasing $t'/t$, both the DOS at VHS increases and the bandwidth decreases (see appendix B3), and thus the NFL phenomena also increases. It is already known that the optimal $T_c$ increases with increasing $t^{\prime}/t$,\cite{Pavarini} and with decreasing $n$. This phenomena is consistently observed in various cuprates, pnictides and heavy-fermions.\cite{twodomes} Our results thus give a microscopic explanation into this empirical observation.

\section{Discussions}\label{Sec:Discussion}

\subsection{Analytic self-energy in the NFL state}\label{Sec:Analytic}

One of the important properties of the present results is that the self-energy is free from any essential singularity and non-analytic form at all momenta, energy, and doping. From Eq.~\eqref{selfenergy}, we can deduce that the self-energy can become non-analytic when either the potential $V_{\nu}({\bf k},\epsilon)$ or the spectral function $A({\bf k},\omega)$, has a non-analytic form. Both these cases are discussed separately below.

(a) Near a Hertz-Millis QCP, there arises a singularity in the spin and/or charge potential $V_{\nu}$ at a characteristic wavelength, causing {\it massless} magnons or plasmons, respectively. Here we focus on the near-optimal doping region which is far away from the AF and CDW QCPs. And as discussed in the main text, paramagnons remain massive at all momenta and doping, and gives no singular behavior. So, $V_{\nu}$ has no essential singularity in the doping range of present interest. Yet we can make few remarks. An AF QCP induced NFL model have been used earlier by Moriya {\it et al.}\cite{MoriyaUeda}. They found that the $T$-linear behavior in resistivity and $d$-wave superconductivity both arise from the strong AF fluctuations.\cite{KontaniReview} If this result holds in cuprates, one would obtain a $T$-linear NFL state at 5-7\% doping. But the $T$-linear behavior is rather shifted to the optimal doping, where the AF fluctuations are negligibly small.\cite{twodomes,MarcJulien} The model was extended by Monthoux and Pines,\cite{NAFL} Millis-Monien-Pines\cite{PinesMillis} with a phenomenological model of the spin-fluctuation. Bicker {\it et al.} used a similar self-consistent FLEX model\cite{FLEX} of the spin-fluctuation mediated NFL calculations. But in all these models, the driving instability has been the the same ${\bf Q}=(\pi,\pi)$ AF fluctuation, and thus the realistic region of NFL state should be 5-7\% doping. In a fully self-consistent scheme, the spin-fluctuation spectrum is modified by the self-energy effect, and such a renormalization effect is sometimes distinguished as the `mode-mode coupling' effect.\cite{Moriyabook} In the mode-mode coupling theory, the magnetic instability is clearly modified, or sometimes removed due to the suppression of the spin-susceptibility from the self-energy correction. As a result, the long-range AF order does not occur in pure 2D systems, which means that the Mermin–Wagner theorem is satisfied here. In reality, the hole-doped cuprates exhibit an AF critical point around 5-7\% doping {\it without} any apparent $T$-linear resistivity.\cite{Taillefer,HusseyScience,twodomes,Kivelson} There can be various reasons, such as finite three-dimensionality in cuprates,\cite{kzdispersion} second-order vertex correction (AL term),\cite{Hartnollop,Tremblayop} non-perturbative corrections,\cite{SSLeeReview} etc., but it is not the main topic of our present work.

(b) Another possible source of singularity is the VHS in the single-particle spectral function $A({\bf k},\omega)$. An earlier DMFT calculation in a single band Hubbard model showed that as the VHS is positioned exactly at the Fermi level, it gives rise to a non-analytical self-energy and thus one cannot treat the transport relaxation rate coming from the single-particle broadening.\cite{cuspVHS} Such a singularity is removed in our case due to multiple reasons and we obtain analytical self-energies even at the extreme NFL region. To understand this, we can write the imaginary part the of self-energy in an  approximate from (from Eq.~\eqref{selfenergy}) as
\begin{eqnarray}
\Sigma_{\nu}^{\prime\prime}({\bf k},\omega)\propto \sum_{\bf q}\int d\epsilon V_{\nu}({\bf q},\epsilon)A({\bf k}-{\bf q},\omega+\epsilon).
\label{ImSigma}
\end{eqnarray} 
In a local approximation where the potential is replaced with a ${\bf q}$-averaged potential, the analyticity of the self-energy is solely determined by the analyticity of the VHS. Therefore, if the VHS has the non-analytic cusp even after including the self-energy correction, the self-energy also becomes non-analytic. 

When the ${\bf k}$-dependent self-energy is introduced, we can see in another way that the VHS is substantially weakened. Near the VHS region around $k_{\rm v}=(\pi,0)$, the first $k$-derivative of the bare dispersion vanishes, and thus the leading term in the band is $\xi_{\bf k}\approx k^2/m^*$, where $k$ is measured with respect to $k_{\rm v}$ ($\hbar=1$). Since $\xi_{\bf k}$ is a slowly varying function in momentum, one obtains a `flat-band', leading to a non-analytic cusp in $d\ge 2$, and a logarithmic divergence in $d=1$. In the ${\bf k}$-dependent self-energy correction, the renormalized band obtains an effective $k$-linear term from the self-energy as $\bar{\xi}_{\bf k}\approx \bm{\nabla} \Sigma^{\prime}\cdot {\bf k} + (1/m^* + \nabla^2 \Sigma^{\prime})k^2$, where the $k-$derivatives are taken at ${\bf k}_{\rm V}$. This linear-in-$k$ terms effectively destroys the essential criterion for a singularity at the VHS. 

\subsection{Sum rules and Luttinger theorem}\label{Sec:Sumrules}
Luttinger theory remains valid with the self-energy correction. This can be easily seen by the fact that $\Sigma''({\bf k},0)=0$ at all momenta. The spectral function obtains isolated poles on the FS at $\bar{\xi}_{{\bf k}_F} = \xi_{{\bf k}_F}-\bar{\mu}+\Sigma'({\bf k}_{F},0)$, where $\xi_{\bf k}$ is understood to be the non-interacting dispersion without the chemical potential. We note that the chemical potential $\bar{\mu}$ is different from that without the self-energy correction. When the self-energy is included, the chemical potential is adjusted to keep the number of electron conserved. 

The $f$-sum rule in the spin and charge channels are also individually satisfied. This can be proven in two ways. The vertex correction is important in the self-consistent scheme and usage of the Ward identity in the vertex correction ensures that the sum-rules remain intact. The basic principle in maintaining the sum rule is that one invokes the similar approximation in both density-, current-correlations functions as well as in the vertex function, and make sure that the Ward identity is followed. The $f$-sum rule for the densities\cite{Tremblayop} is 
\begin{eqnarray}
&&\frac{1}{\pi}\int d\epsilon\epsilon \Gamma_{\nu}({\bf q},\epsilon)V_{\nu}({\bf q},\epsilon) \nonumber\\
&&\qquad = \frac{1}{N}\sum_{{\bf k}}(\xi_{{\bf k}+{\bf q}} -\xi_{{\bf k}-{\bf q}}-2\xi_{{\bf k}})\langle n_{\uparrow}\pm n_{\downarrow}\rangle.
\label{fsum1}
\end{eqnarray}
$\pm$ signs indicate charge ($\nu=1$) and spin ($\nu=2$) densities. Since the spin is conserved here, $\frac{1}{\pi}\int d\epsilon\epsilon V_{2}({\bf q},\epsilon)$ must vanish. In the mean-field level without the self-energy correction, the potential $V^0_{\nu}$ satisfy Eq.~\eqref{fsum1}. Let us assume $\bar{V}_{\nu}(q,\epsilon)$ is the $Z$-renormalized potential which is obtained from Eqs.~\eqref{Vq}-\eqref{chi} by replacing the spectral function with its quasiparticle form $A({\bf k},\omega)=Z/(\omega-\bar{\xi}_{\bf k})$. This gives $V_{\nu}({\bf q},\epsilon)\approx Z\bar{V}_{\nu}({\bf q},\epsilon)$. Then we can easily show that the energy range (=bandwidth $\mathcal{W}$) of $\bar{V}$ is reduced by $Z$ (since the band is renormalized by the same $Z$). Since the vertex correction is $\Gamma\sim 1/Z$, we obtain $\Gamma({\bf q},\epsilon)V_{\nu}({\bf q},\epsilon)\approx V^0_{\nu}({\bf q},\epsilon)$. This is a direct consequence of the Ward identity in which the kinetic energy and the interaction potential are renormalized by the same factor $Z$, and thus the intermediate coupling scenario remains valid with and without including the self-energy correction. 

Similarly, we can prove that the optical sum rule also remains valid here. As mentioned in Sec.~\ref{Sec:Resistivity}, the momentum dependent self-energy leads to a current-current vertex correction ${\bf \Gamma}$ which arises from the ${\bf k}$-derivative of the self-energy\cite{def_sus}. The current vertex is again related to the density vertex $\Gamma$ via the Ward identity. The optical conductivity in terms of the Matsubara frequency, in the limit of ${\bf q}\rightarrow 0$, can be written as
\begin{eqnarray}
\sigma(i\epsilon_m) &=& e^2\frac{1}{N}\frac{1}{\beta}\sum_{{\bf k},n}{\bf v}_{k}\cdot{\bm \Gamma}({\bf k},i\omega_n,i\epsilon_m)\nonumber\\
&&\times G({\bf k},i\omega_n)G({\bf k},i\omega_n+i\epsilon_m).
\label{currentsum1}
\end{eqnarray}
Now from the Ward identity (see Eq.~\eqref{ward}), we substitute $m{\bf v}_{k}\cdot{\bm \Gamma}({\bf k},i\omega_n,i\epsilon_m)=G^{-1}({\bf k},i\omega_n)-G^{-1}({\bf k},i\omega_n+i\epsilon_m) +i\epsilon_n\Gamma({\bf k},i\omega_n,i\epsilon_m)$, where $\Gamma({\bf k},i\omega_n,i\epsilon_m)$ is the density vertex. We get
\begin{eqnarray}
\sigma(i\epsilon_m) &=& \frac{e^2}{m}\frac{1}{N}\frac{1}{\beta}\sum_{{\bf k},n}\left[G({\bf k},i\omega_n+i\epsilon_m)-G({\bf k},i\omega_n)\right. \nonumber\\
&& \left. + i\epsilon_n\Gamma({\bf k},i\omega_n,i\epsilon_m)G({\bf k},i\omega_n)G({\bf k},i\omega_n+i\epsilon_m)\right].\nonumber\\
\label{currentsum2}
\end{eqnarray}
In a homogeneous charge medium, the first two terms cancel each other. The last term $\frac{1}{N}\frac{1}{\beta}\sum_{{\bf k},n}\left[\Gamma({\bf k},i\omega_n,i\epsilon_m)G({\bf k},i\omega_n)G({\bf k},i\omega_n+i\epsilon_m)\right]$ is bare charge density susceptibility $\chi({\bf q}\rightarrow 0,i\epsilon_m)$. Now from the $f$-sum rule for density in Eq.~\eqref{fsum1} we get $\frac{1}{\beta}\sum_m i\epsilon_m \chi({\bf q}\rightarrow 0,i\epsilon_m)=\pi n/2$, where $n$ is the total charge density. Therefore, we get $\frac{1}{\beta}\sum_m \sigma (i\epsilon_m) = \frac{\pi ne^2}{2m}=\omega_{pl}^2/8$, where $\omega_{pl}$ is the plasma frequency. The optical sum rule implies that the total absorbing power of the solid characterized by $\sigma$ does not depend on the details of the interactions and is determined only by the total number of particles in the system.\cite{Pinesbook,opsumrule} Such a sum rule is modified if the FS is partially or fully incoherent,\cite{Abanov_opsumrule} which is not the case in our model.
 
\subsection{Other angular-dependent self-energy calculations}\label{Sec:Angle}
Angle-dependent self-energy and NFL state have been studied earlier in a variety of approaches. Usually in cluster DMFT\cite{CDMFT} and Dynamical Cluster Approximation (DCA)\cite{CQMC}, the momentum dependent calculation is done in small clusters and the results are in general agreement with ours. In FLEX and GW methods, which can retain the full spectrum of the correlation potential, one can account for the full-momentum dependence of the self-energy.\cite{Kontani_FLEX,GW_kdepSE,DasMottAFM,MRDFNickelate} In an earlier FLEX calculation\cite{Kontani_FLEX}, it was found that the self-energy effect is maximum at the AF `hot-spot', rather than at the antinodal points. The apparent discrepancy between the FLEX and our MRDF method arises from how the spin-fluctuation potential is treated. FLEX calculation only included the AF fluctuation, and does not include paramagnons. So, its range of validity is limited below $x<0.10$ where the AF fluctuation is present. Also, in the context of heany-fermion compounds, it was shown that a strongly anisotropic hybridization can generate angular dependent quasiparticle residue.\cite{Ghaemi_kdep} There are also non-perturbative calculations of the angle-dependent NFL state in the strong coupling region.\cite{SSLee_kdep} Their results are in general agreement with the FLEX calculation that the NFL state is stronger at the AF `hot-spot'. Our method includes both AF and paramagnons fluctuations and show strong paramagnon dresssed self-energy effect at the antinodal points in the optimal doping region. Finally, our obtained self-energy anisotropy is in qualitative agreement with a QMC calculation of a single band Hubbard band where the correlation is treated mainly for the paramagnon fluctuations.\cite{JarrellEPL}


\subsection{NFL induced Hertz-Millis QCP}\label{Sec:QCMfromNFL}
As discussed above in various  context, within the self-energy picture, two sources of NFL behavior are primarily discussed; through the singularities in the bosonic spectrum, or through that of the single particle spectral function. A major part of the literature discusses the origin of NFL state from the QCP physics, in which one obtains singularities in the bulk properties due to the singularities in the bosonic spectrum $V_{\nu}({\bf q},\epsilon)$. In another case, mass divergence of the quasiparticle spectrum $A({\bf k},\omega)$ can introduce non-analytic self-energy. A related situation arises in the case of a Pomeranchuk instability due to `soft' FS, which gives strongly enhanced decay rate for single-particle excitations and NFL behavior.\cite{MetznerRohe} More such cases are reviewed by L\"ohneysen {\it et al.} (in Sec.~IIIG of Ref.~\onlinecite{PWolfle}). Here, we obtain a different model where the dynamical itinerant-local density fluctuation causes the NFL behavior only in certain parts of the BZ, and it adiabatically connects to the FL region with analytic self-energy. So, we can ask a question: can the NFL state (without the QCP origin) give a QCP? Mermin-Wagner theorem prohibits the order  induced by density fluctuations in two-dimensions. In the mode-mode coupling theory,\cite{Moriyabook,MoriyaUeda} it is shown for a AF fluctuation that the self-energy reduces the spectral weight at the magnetic `hot-spot' and thereby weakens the static nesting. Therefore, NFL state would oppose the formation of a QCP. According to the Hertz-Millis theory\cite{HertzMillis} both dynamical and static fluctuations are related to each other at the QCP.  In our momentum dependent calculation, we find that the anisotropic self-energy is actually a nonlocal effect (see Sec.~\ref{Sec:Selfenergy}). What we mean by this is that the dominant self-energy values at the antinodal point are mainly contributed by the incoherent, high-energy Hubbard bands at the BZ center and corner [$\Gamma$, ($\pi,\pi$)]. Therefore, the states away from the NFL momenta [$(\pi,0)/(0,\pi)$] can develop static orders if a suitable FS nesting is present. As in the case of cuprates, the NFL state at the optimal doping resides at the antinodal point, while the AF state and the $d$-wave superconductivity arise from the FS nesting at the magnetic `hot-spot' (within a weak/intermediate coupling scenario). In fact, as the spectral weight is transferred from the antinodal to the rest of the BZ, the magnetic `hot-spots' gain more spectral weight and the corresponding nesting can be enhanced. The present NFL state will however disfavor the charge density wave (CDW) which is believed to arise from the antinodal nesting.\cite{Kivelson} Our prior calculation indeed showed that the CDW nesting is shifted from the antinodal region to the tip of the `Fermi arc' below the magnetic BZ, which is consistent with experiments.\cite{Fermiarc} However, such a CDW is also predicted to give a discontinuous, first-order phase transition near the optimal doping to avoid the nesting at the antinodal point.\cite{Fermiarc}

\subsection{Pseudogap}\label{Sec:Pseudogap}
The discussion of a pseudogap feature follows from the above section. In the present model, there is a `Fermi arc' due to strong suppression of the spectral weight at the antinodal points, see Fig.~\ref{fig3}. However, the entire `Fermi arc' remains coherent. In the angle-integrated density of states, no suppression of the spectral weight is obtained at the Fermi level. In other words, the `Fermi arc' does not produce a pseudogap in the DOS. The doping dependence of the `Fermi arc' is discussed in a separate work.\cite{Fermiarc} There is an increasing discussion that the pseudogap originates from some  sort of a competing order, whose origin is yet to be determined. Any competing order induced gap in the low-energy state may not affect much the NFL state. This is because the pseudogap is typically of the order of 50-80meV, while the itinerant-local density fluctuations energy is 300-500meV even at the optimal doping. Therefore, we expect that the pseudogap will have less influence on the NFL physics. Experimentally, the resistivity-$T$ exponent is derived above the pseudogap temperature $T^*$. In our calculations also we have reported the exponent for $T> 100$ K in Fig.~\ref{fig4}. 

\subsection{NFL to FL with decreasing $U$}\label{Sec:U}
We have mentioned before that the Hubbard $U$ determines the overall strength of the NFL state, but not the $k-$space anisotropy. Again, we discussed in the resistivity calculation that the global bulk NFL/FL property of the system is determined by how much $k$-space volume each self-energy occupies for a given value of $U$. This leads to a question: how does one obtain the FL behavior by continuously reducing the value of $U$? 

In the Appendix ~\ref{Apx:U}, we have repeated all the results for different values of $U$. We indeed find that the momentum profile of $p_k$, doping dependence of $n$ etc. remain the same for different values of $U$. However, their overall strength decreases with decreasing $U$. We also find that as we reach the weak-coupling regime, where the fluctuations become irrelevant, the self-energy still remains equally anisotropic, but the range of variation of $p_k$ narrows down to be around $2$ only. This gives the resistivity-$T$ exponent $n\rightarrow 2$. Thus the FL state is recovered in the weak-coupling region (see appendix B). 

\section{Conclusions}\label{Sec:Conclusion}
The important message of our result is that for strongly anisotropic materials where the dynamical fluctuations have significant momentum dependence, the resistivity-temperature exponent is not a robust measure of the full correlation spectrum of the underlying quasiparticle states. We found that even in the underdoped and overdoped regions, where resistivity exponent $n\rightarrow 2$, there are considerable amount of NFL self-energies lying in the antinodal regions. Similarly, in the extreme NFL region near the optimal doping regime (determined by $n\sim 1$), the nodal quasiparticles continue to behave FL-like (with $\Sigma^{\prime\prime}\propto|\omega|^2$). Both as a function of temperature and doping (and other tunnings), the spectral weight is transfered between the NFL and FL regions and the system adiabatically transforms from a dominant NFL to a FL-like state, as seen in experiments. Our work suggests that the microscopic and macroscopic landscapes of the NFL behavior can be characteristically different and that a direct correspondence between $k$-resolved spectroscopy (such as ARPES, and quasiparticle interference (QPI) pattern) and the transport, and thermodynamical properties are necessary to deduce the global and local NFL behavior of a given system.

\noindent
\vskip0.5cm
\noindent
%
\begin{appendix}

\section{Tight-binding parameters}\label{Apx:TB}
\begin{table}[h!]\label{table}
\begin{tabular}{|m{1.2cm}|m{1.0cm}|m{1.0cm}|m{1.0cm}|m{1.0cm}|m{0.6cm}|m{1.4cm}|}
\hline
Material &  $t$ & $t'$ & $t''$ & $t'''$ & U & Ref.\\
\hline
LSCO & 0.4195 & -0.0375 & 0.018 & 0.034 & 1.6 & \onlinecite{kzdispersion} \\
\hline
YBCO & 0.35 & -0.06 & 0.035 & -0.005 & 1.9 & \onlinecite{DasAIP} \\
\hline
\end{tabular}
\vspace{12pt}\\
\caption{Full tight binding parameters for different materials. All energies are given in eV.}
\label{TableA1}
\end{table}

\section{$U$ dependence of various results}\label{Apx:U}
All results and conclusions presented above are obtained for material specific values of the Hubbard $U$ (see Table~\ref{TableA1}). Here, we investigate them for different values of $U$ and study their evolution. The following results also demonstrate the distinction between the doping dependence of the static correlation ($U$) and the dynamical correlation ($V(\omega)$) in Eq.~\ref{V_pot}).  


\begin{figure}[h!]
\centering
\rotatebox[origin=c]{0}{\includegraphics[width=0.99\columnwidth]{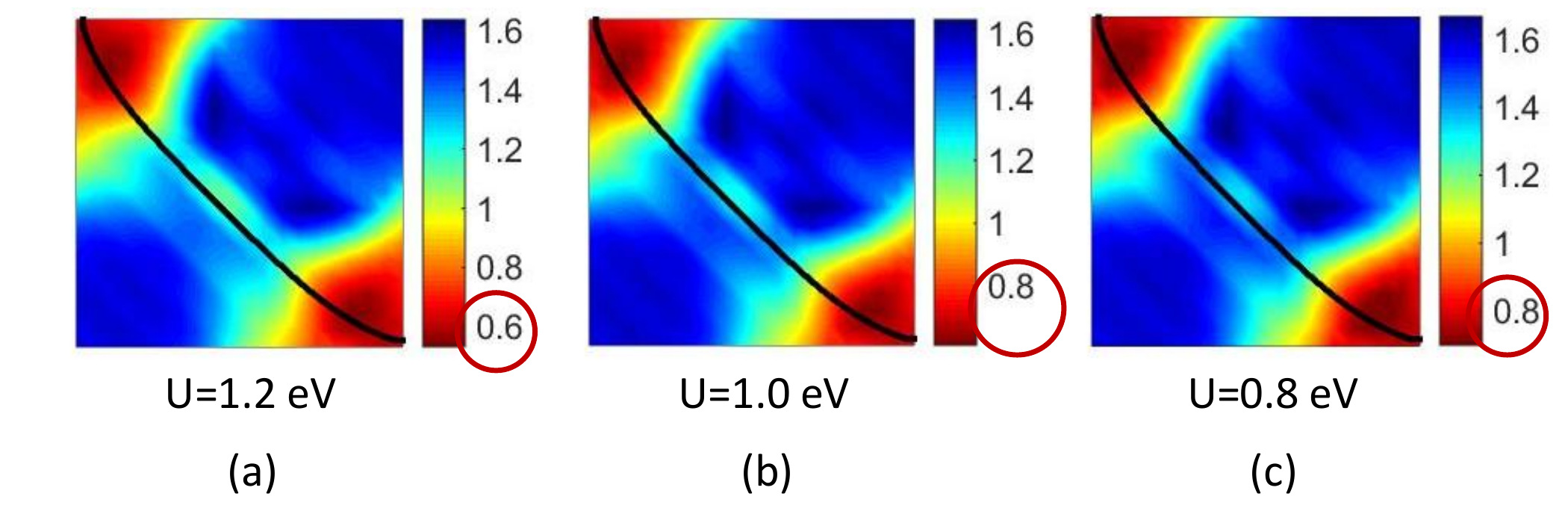}}
\rotatebox[origin=c]{0}{\includegraphics[width=0.80\columnwidth]{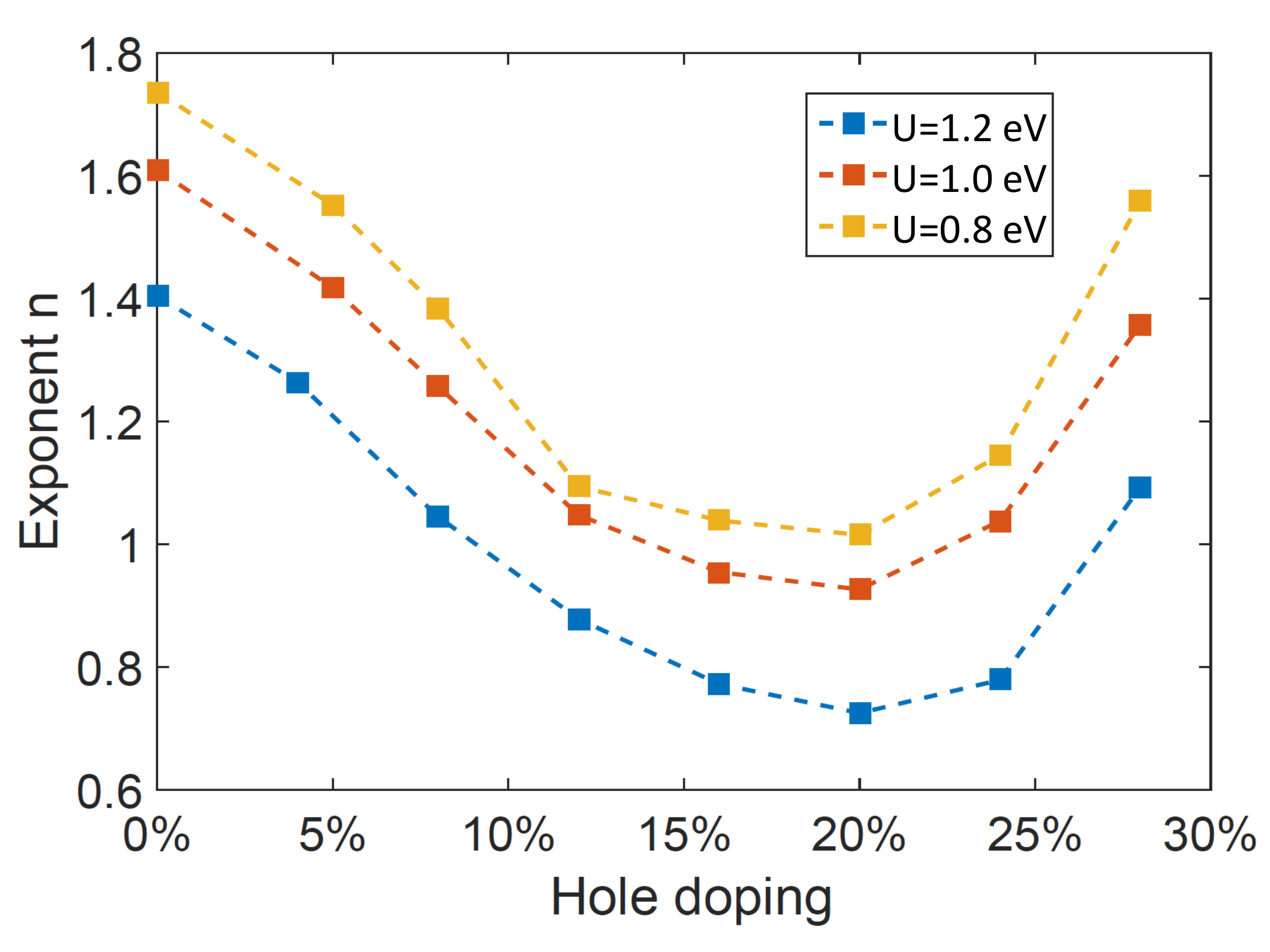}}
\vspace{12pt}\\
\caption{(a-c) Plots of the self-energy -frequency exponent $p_k$ (defined in Eq.~1) for three different values of $U$ for LSCO at $x=0.20$. In all three cases, we notice that the overall momentum profile of $p_k$ remains very much the same. This is expected since the anisotropy is related to the anisotropy in the electronic structure and correlation function, but not directly on the onsite $U$. The overall range of $p_k$ (seen in the adjacent colorbars) however decreases with decreasing $U$. This means the system moves towards the FL state at a fixed doping as $U$ decreases. (Lower panel) The resistivity exponent $n$ is plotted as a function of doping for the same three values of $U$ }
\label{figB1}
\end{figure}

Keeping all other parameters the same, we expect that the system would tend to transform from NFL to FL like as we decrease the values of $U$. This is what we observe in Fig.~\ref{figB1} where we plot the momentum profile of $p_k$ at a fixed doping of $x=0.20$ for LSCO for three different values of $U$. In all three cases, the momentum profile remains very much the same, as we expect, since the momentum dependence is governed by the anisotropy in the band structure and correlation function. We notice a characteristic change in the overall range of $p_k$ (as highlighted by red circles in the adjacent colorbars). We find that both the minimum and the maximum values of $p_k$ increases with decreasing $U$. In addition, we also notice that the $k$-space area of the NFL region ($p_k\sim 1$) also decreases with decreasing $U$, reflecting that the system moves towards FL as correlation weakens. The result is confirmed by the resistivity exponent calculation as presented in the lower panel in Fig.~\ref{figB1}.   



We obtain the same conclusion in the resistivity-temperature exponent $n$, calculated with the same parameter sets as in Fig.~\ref{figB1}. We find that the overall doping dependence of $n$ is similar for all three values of $U$: it obtains the minimum value near the optimal doping where the VHS passes through the Fermi level, irrespective of the values of $U$. However, the overall value of $n$ increases with decreasing $U$ as the system moves towards the `global' FL state with lowering its correlation strength.


\begin{figure}[h!]
\centering
\rotatebox[origin=c]{0}{\includegraphics[width=0.70\columnwidth]{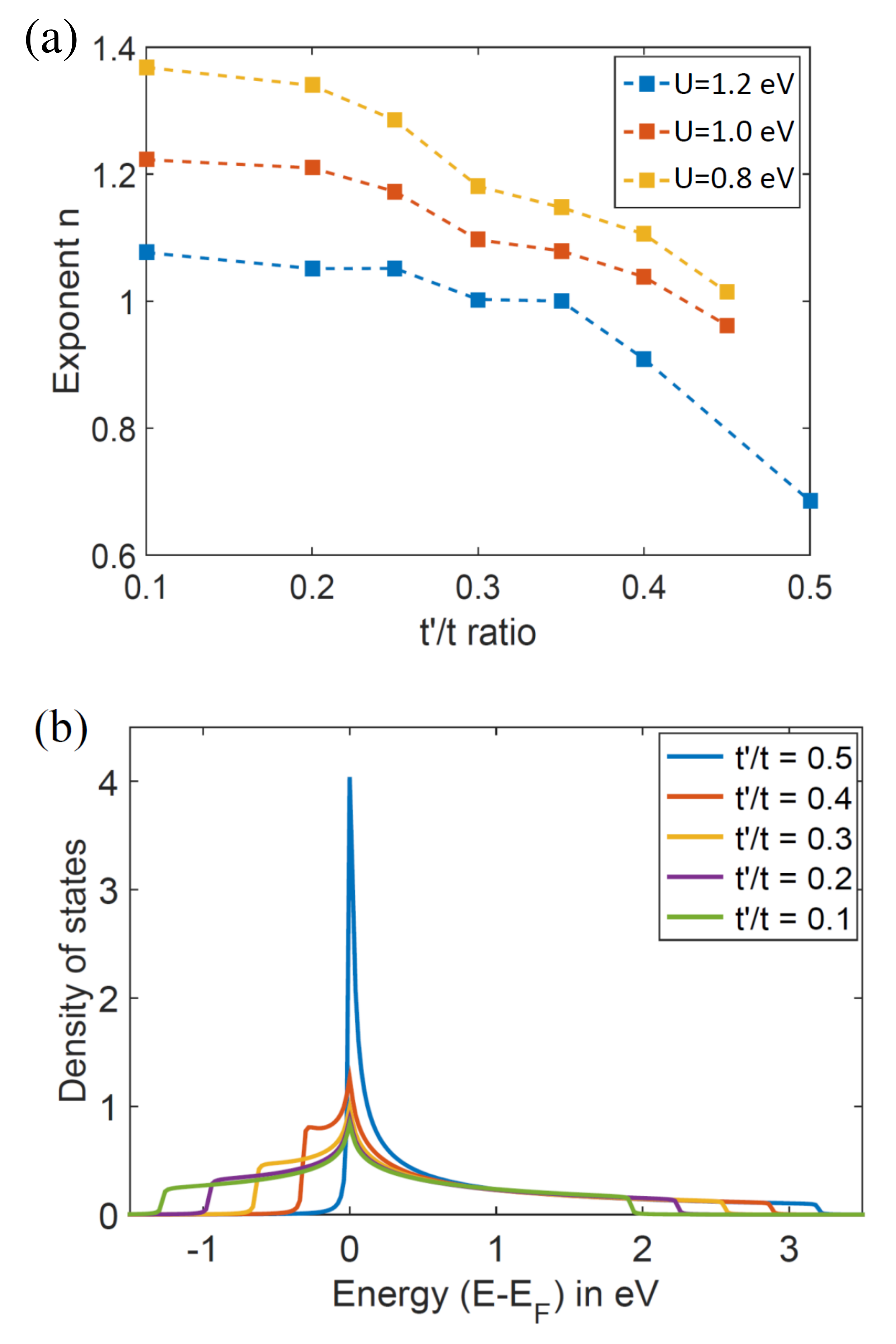}}
\vspace{12pt}\\
\caption{(a) Plots of $t'/t$ vs. $n$ for different values of $U$. As expected, the exponent $n$ decreases with increasing $U$, but for all values of $U$, the $t'/t$ dependence on $n$ is maintained. (b) Density of states (DOS) is shown as a function of energy for different values of the $t'/t$ ratio. Note that the DOS at the VHS gradually increases with increasing $t'/t$ ratio, as the flatness of the band increases at the antinodal point. The Fermi level for all cases is fixed at the VHS.}
\label{figB3}
\end{figure}

Finally we study the evolution of the  $t'/t$ vs. $n$ plot for different values of $U$ in Fig.~\ref{figB3}. We learned in the main text that $n$ decreases as the $t'/t$ ratio increases, keeping the corresponding VHS  fixed at the Fermi level for all cases. This is because the DOS at the VHS increases with increasing $t'/t$ and the bandwidth simultaneously decreases. Therefore, the system becomes more NFL-like as $t'/t$ increases. This conclusion remains intact as we tune the values of $U$. For different values of $U$, the general trend of $t'/t$ vs. $n$ remains the same, however the overall range of $n$ increases with decreasing $U$ as we also found in Fig.~\ref{figB1}.

\section{Extraction of the exponents}\label{Apx:SEextraction}

\begin{figure}[h!]
\centering
\rotatebox[origin=c]{0}{\includegraphics[width=0.99\columnwidth]{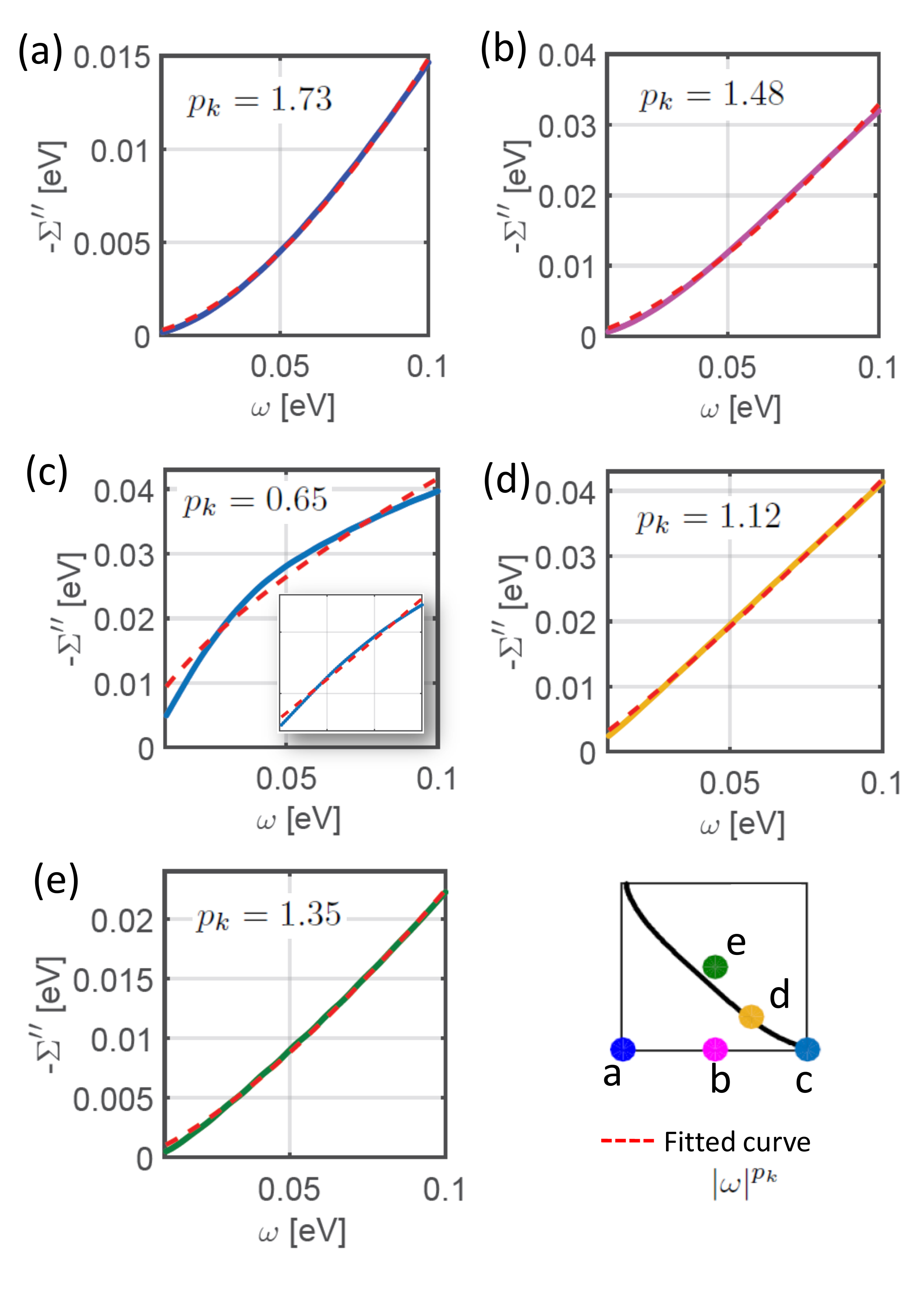}}
\vspace{12pt}\\
\caption{Here we illustrate the curve fitting procedure. In (a-e) we plot $\Sigma^{\prime \prime}({\bf k},\omega)$ (solid line) and fitted curve (dashed line) at different points of BZ as given in Fig.~\ref{fig2}. Same color scheme has been used as indicated in bottom right corner. The red dashed line is the fitted curve with a ${\bf k}$-dependent exponent $p_{k}$ upto a frequency limit $\omega_{u}$. Note that in (c) at $(\pi,0)$-point, the fitting is poor for $\omega_{u}=0.1 eV$, but if we decrease $\omega_{u}$ to 0.04~eV we get a better fitting as indicated in the inset.
}\label{figB4}
\end{figure}

The broadening $\delta$ enters into all quantities and thus modifies the self-energy and resistivity in a complicated way. In the self-energy calculation we use Green's function as $G({\bf k},\omega)=(\omega-\epsilon_{{\bf k}}-\Sigma({\bf k},\omega)+i\delta)^{-1}$ where $\delta$ is the impurity broadening. Without an impurity broadening ($\delta=0$), $\Sigma^{\prime \prime}({\bf k},0)=0$ which causes problem in the Green's function formalism since it has poles on the real axis. On the other hand, finite value of $\delta$ shows up as $\Sigma^{\prime \prime}({\bf k},0)\approx\delta$ in the self-consistent calculations of self-energy. This finite value in turn modifies the intrinsic $\omega$ dependence at low frequency, up to $\sim 2\delta$. So, to extract the correct exponent $p_k$, we exclude the low frequency region of order $2\delta$. 

Furthermore, $p_k$ is also frequency dependent, but usually the frequency dependence is so small at low frequencies that one can approximate it as constant in this frequency range. This behavior is  also observed in experiments where exponent is extracted using an upper limit ($\omega_{u}$) in frequency, and $\omega_{u}$ is found to vary over the BZ.\cite{Cuprate_ARPES} In our calculations, we use a fixed $\omega_{u}$ (0.1 eV) since finding $\omega_{u}$ for every points of the BZ can be ambiguous. We then take the average of exponent in that frequency window. We further illustrate the procedure by plotting the calculated and the fitted curves in Fig.~\ref{figB4}. At the $(\pi,0)$ point, a fixed power law behavior can be obtained only up to a small frequency limit upto 0.04 eV (see inset of Fig.~\ref{figB3}(c)) which is consistent with experimental data.\cite{Cuprate_ARPES}. 2

\begin{figure}[h!]
\centering
\rotatebox[origin=c]{0}{\includegraphics[width=0.95\columnwidth]{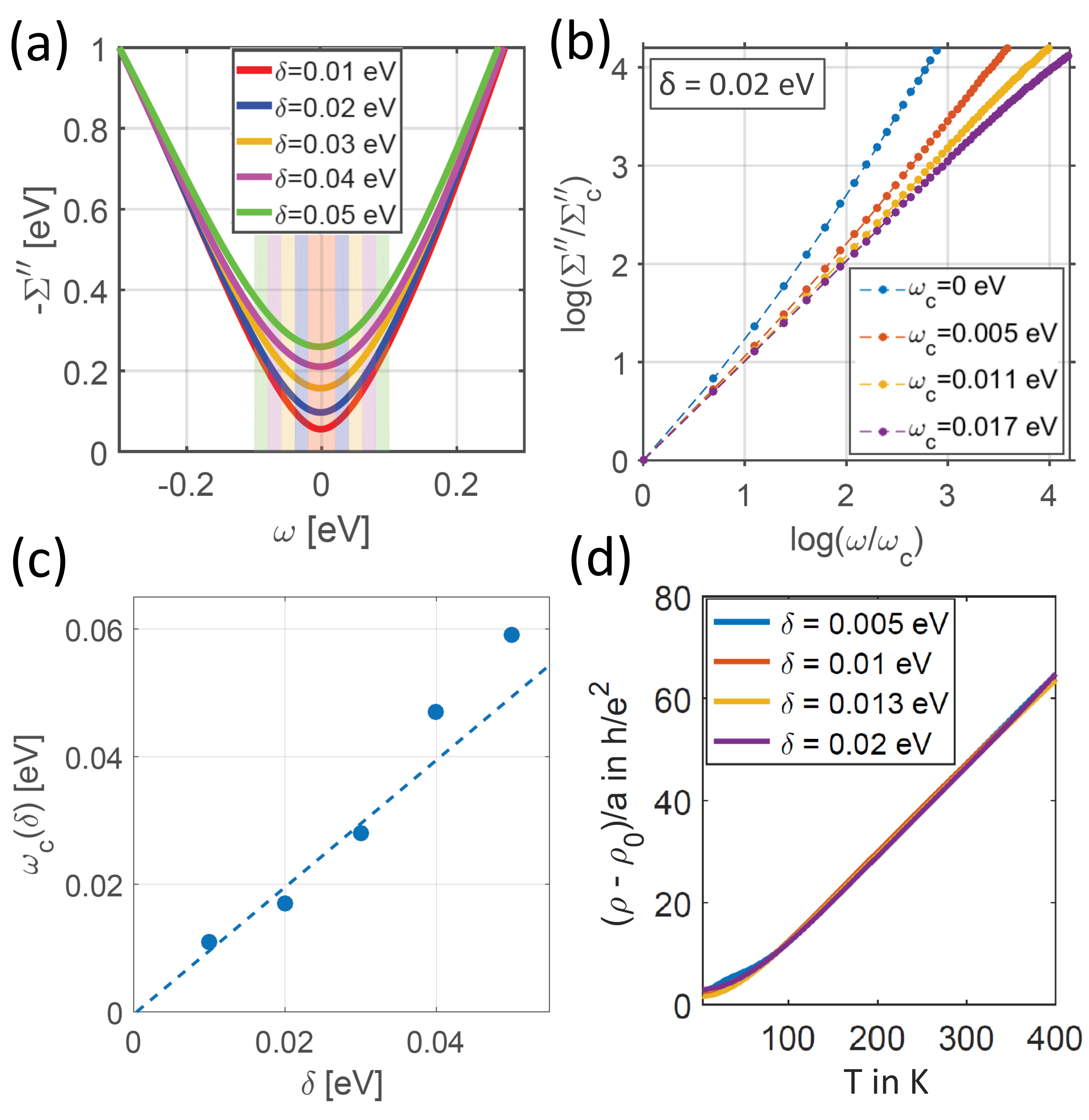}}
\vspace{12pt}\\
\caption{(Color online) Plots of (a) $\Sigma^{\prime \prime}({\bf k},\omega)$ at a sample point, $(0,0)$ point, for different $\delta$ for x=0.20 LSCO, (b) logaithmic plot of $\Sigma^{\prime\prime}$ vs $\omega$ for $\delta=0.02 eV$ with  cut-off frequency $(\omega_{c}(\delta))$ at $(\pi,0)$ point (Only low $\omega$ region is shown in the plot), (c) frequency cut-off $\omega_{c}$ as a function of $\delta$, and (d) resistivity $\rho$ for different $\delta$. Note that with the increase of $\delta$, the low energy region of $\Sigma^{\prime \prime}({\bf k},\omega)$ (shaded area in (a)) becomes flatter, and as a consequence, the low temperature region of $\rho$ also becomes flatter. But at a relatively higher energy or temperature range, the overall power law behavior of $\Sigma^{\prime \prime}({\bf k},\omega)$ as well as $\rho$ remains the same.
}\label{figB5}
\end{figure}

In Fig.~\ref{figB5} (a-d), we further illustrate the effect of $\delta$ on $\Sigma^{\prime \prime}$ and $\rho$. To show that the exponent $n$ is essentially $\delta$-independent, we fit the resistivity curves as $\rho=\rho_{0}+aT^{n}$ by allowing $\rho_{0}$ and $a$ to be $\delta$-dependent [Fig.~\ref{figB5}(d)]. We find that different $\rho(T)$ curves obtained for different broadening values are overlaid on each other, suggesting the exponent $n$ is independent on the choice of the impurity broadening. To clarify the effect of numerical broadening we analyze the logarithmic behavior of $\Sigma''$ for different broadenings. For every broadening, we take a cut-off frequency $\omega_{c}$, and the self-energy data above $\omega_c$ are considered for extracting the frequency exponent. The reason is that below $\omega_c$, the result is influenced by the choice of broadening parameter $\delta$, but above $\omega_c$, the results are independent of the choice of $\delta$.  As expected when this cut-off is zero, the frequency exponent indicates a value greater than one, see Fig. Fig.~\ref{figB5} (b).  If we increase this cut-off we approach the linear behavior and thus we can extract the minimum frequency cut-off that gives linear behaviour. In Fig.~\ref{figB5} (c) we plot this minimum cut-off ($\omega_{c}$) as a function of broadening and thus we are able to show as $\delta$ approaches zero $\omega_{c}$ indeed approaches zero.
\section{Details of the MRDF calculations}\label{Apx:Calculation}

\subsection{Self-energy dressed susceptibilities}

We start with the standard definition of spin/charge susceptibility \cite{def_sus,def_sus1} which is given by
\begin{eqnarray}
\chi^{ij}({\bf q},\tau )  = \dfrac{1}{N}\langle T_{\tau}\Pi^{i}({\bf q},\tau)\Pi^{j}(-{\bf q},0) \rangle,
\label{Eq:chi1}
\end{eqnarray}
where $\Pi^{i}({\bf q},\tau)$ denotes the spin/charge density where indices $i$, $j$ denote different components (for example x, y, z components) in case of spin susceptibility.
Charge and spin densities $\left( \Pi^{i}({\bf q},\tau)\right)$ are given in the second quantized notation as
\begin{eqnarray}
\rho_{{\bf q}}(\tau) =\sum\limits_{{\bf k},\sigma}c^{\dagger}_{{\bf k}+{\bf q},\sigma}(\tau)c_{{\bf k},\sigma}(\tau), \\
S^{i}_{{\bf q}}(\tau) =\sum\limits_{{\bf k},\alpha,\beta} c^{\dagger}_{{\bf k}+{\bf q},\alpha} (\tau)\sigma^{i}_{\alpha\beta}c_{{\bf k},\beta}(\tau),
\end{eqnarray}
where $\sigma^{i}$s are the Pauli spin matrices in 2D. $c^{\dagger}_{{\bf k},\sigma}(\tau)$ is the dressed quasi-particle creation operator (sometimes called Dyson orbital) at the Bloch momentum ${\bf k}$ and spin $\sigma$. Since the ground state is spinless, both transverse and longitudinal spin-densities, as well as the charge density term yield the same bare susceptibility. In general, we can write Eq.~\eqref{Eq:chi1} as 
\begin{eqnarray}
\chi({\bf q},\tau) &&= \dfrac{1}{N} \sum\limits_{{\bf k},{\bf k'},\sigma,\sigma',\sigma'',\sigma'''} \nonumber\\ 
&&\langle T_{\tau} S(\infty) c^{\dagger}_{{\bf k}+{\bf q},\sigma}(\tau)c_{{\bf k},\sigma^{\prime}}(\tau) 
c^{\dagger}_{{\bf k'}-{\bf q},\sigma''}(0)c_{{\bf k'},\sigma'''}(0)\rangle,\nonumber\\
\label{Eq:chi0}
\end{eqnarray}
where the momentum conservation law is imposed. $S(\infty)$ is the usual S-matrix which arises in the interaction picture.\cite{sus_def,Abrikosovbook} We can decompose the four-field terms into bi-linear terms within the Wick's theorem, and allow the spin-conservation condition for the ground state. We restrict ourselves to the bubble diagrams for the density-density correlations and the density vertex correction contains the ladder diagrams. Furthermore, we include only the RPA terms, with all the bubbles containing the same density vertex term. We are not including the higher order ladder diagrams here, which was derived by MT,\cite{MT} and AL.\cite{AL} These two terms are discussed below for the current-current correlation functions (Sec.~\ref{Apx:Cond}), and one would obtain similar terms for the density-density correlation term. We will show that these terms give negligible contributions in the intermediate coupling range, and we defer its discussion to Sec.~\ref{Apx:Cond} below. Since the ground state has both spin-rotational and gauge symmetry, the bare spin and change susceptibilities are the same without the vertex term. They become decoupled in the RPA label, and give different self-energies for the spin and charge channels. In our self-consistent approximation, the vertex correction depends on the self-energy, and thus it has different contributions from the spin and charge sectors. Therefore, it makes more sense to decouple the bare spin ($\nu=1$) and charge ($\nu=2$) susceptibilities at this bare level, and we obtain 
\begin{eqnarray}
\chi_{\nu}({\bf q},\tau)&=&\dfrac{1}{N}\sum\limits_{{\bf k},{\bf k'},\sigma,\sigma',\sigma'',\sigma'''}\langle T_{\tau} c_{{\bf k},\sigma'}(\tau) c^{\dagger}_{{\bf k'}-{\bf q},\sigma''}(0)\delta_{\sigma'\sigma''}\rangle \nonumber\\
&&\hspace{0.5cm} \times\langle T_{\tau} c_{{\bf k'},\sigma'''}(0)c^{\dagger}_{{\bf k}+{\bf q},\sigma}(\tau)\delta_{\sigma\sigma'''}\rangle\Gamma_{\nu}({\bf k},{\bf k}+{\bf q},\tau), \nonumber\\
&=&\dfrac{1}{N}\sum\limits_{{\bf k},\sigma\sigma'}G_{\sigma'}({\bf k},\tau)G_{\sigma}({\bf k}+{\bf q},-\tau)\Gamma_{\nu}({\bf k},{\bf k+q},\tau),\nonumber\\
\end{eqnarray}
where we have identified the terms in the brackets as self-energy dressed Green's functions.
Using the Fourier transformation, we get 
\begin{eqnarray}
&&\chi_{\nu}({\bf q},i\epsilon_{m})=\dfrac{1}{N\beta^{2}}\sum\limits_{{\bf k},n,n'}\int\limits_{0}^{\beta} d\tau e^{( -i\omega_{n}+i\omega_{n'}+i\epsilon_{m})\tau}\nonumber\\
&&\qquad\times \Gamma_{\nu}({\bf k},{\bf k}+{\bf q},i\omega_n,i\epsilon_m)G({\bf k},i\omega_{n})G({\bf k}+{\bf q},i\omega_{n'}), \nonumber \\
&&\qquad=\dfrac{1}{N\beta}\sum\limits_{{\bf k},i\omega_{n}}\Gamma(k,k+q)G(k)G(k+q)).
\label{Eq:chi2}
\end{eqnarray}
We use compact, four-vector, notation $k=({\bf k},i\omega_n)$, and $k+q=({\bf k}+{\bf q},i\omega_{n}-i\epsilon_{m})$. Here $i\omega_n$ and $i\epsilon_{m}$ are the fermionic and bosonic Matsubara frequencies, respectively. From here onwards we drop the index $\sigma$ and assume an implied sum over $\sigma$ index. It is not easy to perform the Matsubara frequency summation using self-energy dressed Green's function. So we use its spectral representation as
\begin{eqnarray}
G({\bf k},i\omega_{n})=\int\limits_{-\infty}^{\infty} \dfrac{d\omega'}{2\pi} \dfrac{A({\bf k},\omega')}{i\omega_{n}-\omega'},
\label{GtoA}
\end{eqnarray}
where the corresponding spectral weight defined as $A({\bf k},\omega)=-{\rm Im}G({\bf k},\omega)/\pi$, where $G({\bf k},\omega)$ is obtained by taking the analytical continuation to the real frequency $i\omega_n=\omega+i\delta$, with $\delta$ being infinitesimal broadening. So the susceptibility expression becomes
\begin{eqnarray}
\chi_{\nu}(q)&=&\dfrac{1}{N}\sum\limits_{{\bf k}}\int\limits_{-\infty}^{\infty}\int\limits_{-\infty}^{\infty} \dfrac{d\omega_{1}}{2\pi}\dfrac{d\omega_{2}}{2\pi} \nonumber\\
&&\times \Gamma_{\nu}(k,k+q)A({\bf k},\omega_{1})A({\bf k}+{\bf q},\omega_{2})\nonumber\\
&&\times \left(\dfrac{1}{\beta}\sum\limits_{n} \dfrac{1}{i\omega_{n}-\omega_{1}}\dfrac{1}{i\omega_{n} -i\epsilon_{m}-\omega_{2}}\right).
\label{beforeMatSum}
\end{eqnarray}
Consistently, we define $q=({\bf q},i\epsilon_m)$. The term in the bracket can be evaluate by the Matsubara summation technique \cite{def_sus} and we arrive at the expression
\begin{eqnarray}
\chi_{\nu}(q)&=&\dfrac{1}{N}\sum\limits_{{\bf k}}\int\limits_{-\infty}^{\infty}\int\limits_{-\infty}^{\infty} \dfrac{d\omega_{1}}{2\pi}\dfrac{d\omega_{2}}{2\pi} A({\bf k},\omega_{1})A({\bf k}+{\bf q},\omega_{2})\nonumber\\
&&\qquad \times \Gamma(k,k+q)\dfrac{f(\omega_{1})-f(\omega_{2})}{i\epsilon_{m}-\omega_{2}+\omega_{1}},\nonumber\\
\label{afterMatSum}
\end{eqnarray}
where $f(\omega)$ denotes the Fermi distribution function. The computation of the susceptibility is done using analytical continuation to the real frequency as discussed before. The susceptibility in the RPA becomes 
\begin{eqnarray}
\chi^{\rm RPA}_{\nu}(q)=\dfrac{\chi_{\nu}(q)}{1\pm U\chi_{\nu}(q)},
\end{eqnarray}
for charge and spin, respectively.

\subsection{Self-energy}
Next we calculate the self energy using the Hedin's approach,\cite{Hedin} which is given by
\begin{eqnarray}
\Sigma_{\nu}(k)&=&-\dfrac{1}{N\beta}\sum\limits_{{\bf q},m}G(k+q)W_{\nu}(q) \Gamma(k,k+q).
\label{Eq:Se1}
\end{eqnarray}
$W$ is the fluctuation-exchange potential which we obtain within the RPA as $W_{\nu}(q)=\frac{\eta_{\nu}}{2}U^2 \chi^{\rm RPA}_{\nu}(q)$, where $\eta=3$ and 1 for spin ($\nu=1$) and charge ($\nu=2$) density fluctuations. Again, to aid the Matsubara frequency summation, we use the spectral representation of $W$ as
\begin{eqnarray}
W_{\nu}({\bf q},i\epsilon_{m})=\int\limits_{-\infty}^{\infty}\dfrac{d\varepsilon'}{2\pi}\dfrac{{\rm Im}\left[ W_{\nu}({\bf q},\varepsilon')\right]}{i\epsilon_{m}-\varepsilon'}.
\label{Eq:W1}
\end{eqnarray}
We denote the fluctuation-exchange potential as $V_{\nu}({\bf q},\varepsilon)={\rm Im} \left[W_{\nu}({\bf q},\varepsilon)\right]$. Therefore, using Eqs.~\eqref{Eq:Se1} and \eqref{Eq:W1}, we get
\begin{eqnarray}
&&\Sigma_{\nu}(k)\nonumber\\
&&=-\dfrac{1}{N}\sum\limits_{\bf q}\int\limits_{-\infty}^{\infty}\dfrac{d\varepsilon}{2\pi} \int\limits_{-\infty}^{\infty}\dfrac{d\omega'}{2\pi}A({\bf k}-{\bf q},\omega')V_{\nu}({\bf q},\varepsilon) \nonumber\\
&&\hspace{0.3cm}\times \Gamma(k,k+q)\left(\dfrac{1}{\beta}\sum\limits_{\epsilon_{m}} \dfrac{1}{i\omega_{n}-i\epsilon_{m}-\omega'}\dfrac{1}{i\epsilon_{m}-\varepsilon}\right) \nonumber \\
&&=\dfrac{1}{N}\sum\limits_{\bf q}\int\limits_{-\infty}^{\infty}\dfrac{d\varepsilon}{2\pi} \int\limits_{-\infty}^{\infty}\dfrac{d\omega'}{2\pi}A({\bf k}-{\bf q},\omega')V_{\nu}({\bf q},\varepsilon) \nonumber\\
&&\hspace{2.3cm}\times \Gamma(k,k+q)\dfrac{f(-\omega')+n(\varepsilon)}{i\omega_{n}-\omega'-\varepsilon},\\
&&=\dfrac{1}{N}\sum\limits_{\bf q}\int\limits_{0}^{\infty}\dfrac{d\varepsilon}{2\pi} \int\limits_{-\infty}^{\infty}\dfrac{d\omega'}{2\pi}A({\bf k}-{\bf q},\omega')V_{\nu}({\bf q},\varepsilon)\nonumber\\
&&\hspace{0.3cm}\times\Gamma(k,k+q)\left[\dfrac{1-f(\omega')+n(\varepsilon)}{i\omega_{n}-\omega'-\varepsilon}+ \dfrac{f(\omega')+n(\varepsilon)}{i\omega_{n}-\omega'+\varepsilon}\right].\nonumber\\
\label{App:selfenergy}
\end{eqnarray}
All other symbols are defined in the main text.

The MRDF method is very similar to the Hedin's equations of self-energy calculation using density-density fluctuations\cite{Hedin}. Different approximations are usually distinguished by different models, such as FLEX\cite{FLEX} or GW methods\cite{GWwoVertex,QSGW}. In the FLEX approach\cite{FLEX}, one calculates the single-particle green's function self-consistently, but not the two-particle one. The extension of the FLEX method where both the single-, and two-particle terms include self-energy correction in a self-consistent way is called the mode-mode coupling theory.\cite{Moriyabook,MoriyaUeda} While in the GW-approach, one often neglects the vertex correction or use a quasiparticle$-$GW approximation etc\cite{QSGW}. In our MRDF approach, we calculate the single-particle Green's function, the density-density correlation function, and the vertex correction by including the self-energy correction. 

\subsection{Optical conductivity}\label{Apx:Cond}

\begin{figure}[t]\label{Feynmann_diagram_2}
\centering
\rotatebox[origin=c]{0}{\includegraphics[width=0.99\columnwidth]{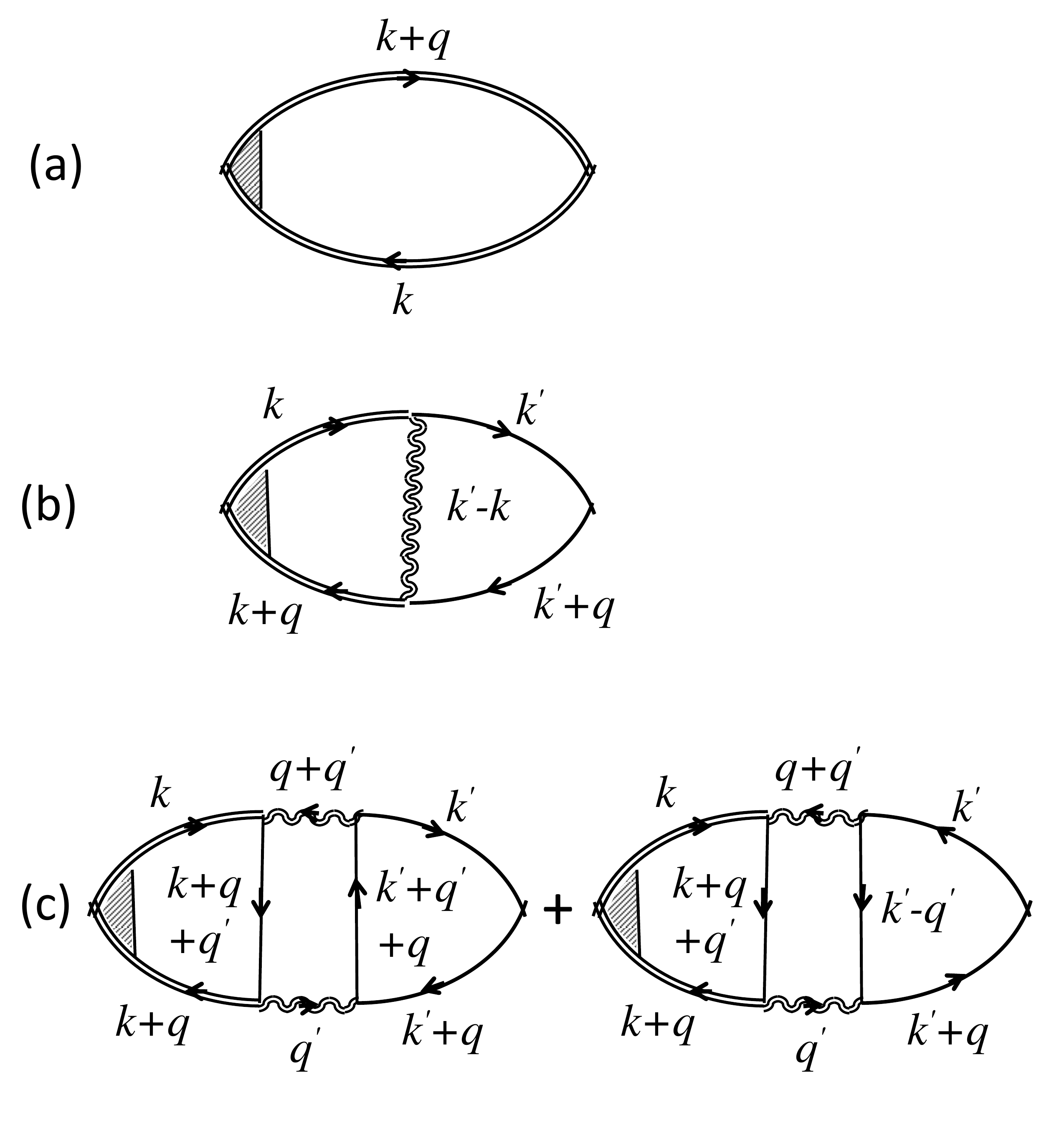}}
\caption{Diagrams of various quantities for conductivity calculation: (a) bubble diagram from Kubo formula (b) Maki-Thomson (MT) diagram, (c) Aslamasov-Larkin (AL) diagram.}
\label{Feynmann_diagram_2nd}
\end{figure}

Kubo formula works well in the weak-coupling region. Maki-Thomson (MT)\cite{MT}, and later Aslamasov-Larkin (AL)\cite{AL} extended the calculations to include higher order diagrams. After deriving them, we will argue below that they can be neglected even in the intermediate coupling region of present interest. In the linear response theory, we have optical conductivity $\sigma_{xx}(\omega)=\frac{1}{\omega}{\rm Im}\mathcal{K}_{xx}({\bf q}\rightarrow 0,\omega)$, where $\mathcal{K}_{xx}$ is the current-current correlation function. (This formula works when $\sigma$, and $\mathcal{K}$ have no singularity). Here $\mathcal{K}_{xx}({\bf q},\tau)=i\langle T_{\tau}[j_x({\bf q},\tau),j_x(-{\bf q},0)]\rangle$, where $j_x({\bf q},\tau)=\sum_{{\bf k},\sigma}v_{x}({\bf k})c_{{\bf k},\sigma}^{\dag}(\tau)c_{{\bf k}+{\bf q},\sigma}(\tau)$ is the current operator. Substituting them, we get
\begin{eqnarray}
\mathcal{K}_{xx}({\bf q},\tau) &&= \dfrac{C}{N} \sum\limits_{{\bf k},{\bf k'},\sigma,\sigma''}\langle T_{\tau} S(\infty) v_{x}({\bf k}) v_{x}({\bf k}') \nonumber\\ 
&&\times c^{\dagger}_{{\bf k}+{\bf q},\sigma}(\tau)c_{{\bf k},\sigma}(\tau) 
c^{\dagger}_{{\bf k'}-{\bf q},\sigma'}(0)c_{{\bf k'},\sigma'}(0)\rangle,\nonumber\\
\end{eqnarray}
The constant factor $C=\frac{e^{2}}{\hbar^{2}}$. Onari {\it et. al.}\cite{CVC}, and Bergeron {\it et al.}\cite{Tremblayop} have derived the explicit for the Kubo, MK and AL terms using diagram approach and the results hold for our MRDF approach. Following the same procedure as in Eqs.~\eqref{Eq:chi0}-\eqref{Eq:chi2}, we can arrive at the first three leading terms. The diagrams for the three terms are given in Fig.~\ref{Feynmann_diagram_2nd}, and the results are 
\begin{eqnarray}
\mathcal{K}^{\rm Kubo}({\bf q},\epsilon_m) &=&\dfrac{C}{3N\beta}\sum\limits_{k}{\bf v}({\bf k})\cdot{\bf \Gamma}(k,k+q)G(k)G(k+q),\nonumber\\
\\
\label{kubo}
\mathcal{K}^{\rm MT}({\bf q},\epsilon_m) &=&\dfrac{C}{(N\beta)^2}\sum\limits_{k,k'}{\bf v}({\bf k})\cdot{\bf \Gamma}(k',k'+q)G(k)G(k+q)\nonumber\\
&&\qquad \times G^0(k')G^0(k'+q)V(k'-k),\\
\label{MK}
\mathcal{K}^{\rm AL}({\bf q},\epsilon_m) &=&\dfrac{C}{(N\beta)^3}\sum\limits_{k,k',q'}{\bf v}({\bf k})\cdot{\bf \Gamma}(k',k'+q')\nonumber\\
&&\qquad\times G(k)G(k+q)G^0(k')G^0(k'+q)\nonumber\\
&&\qquad\times [G^0(k'+q'+q)+G^0(k'-q')]\nonumber\\
&&\qquad\times G^0(k+q'+q)V^{(2)}(q',q'+q).
\label{AL}
\end{eqnarray}
We continue to use the compact notation $k=({\bf k},i\omega_n)$, and $k+q=({\bf k}+{\bf q},i\omega_{n}-i\epsilon_{m})$. $V=V_{1}+V_{2}$ (spin+charge) is the total density fluctuation, and $V^{(2)}(q',q'+q)=V_1(q')V_1(q'+q) +V_2(q')V_2(q'+q)$. ${\bf \Gamma}(k,k+q)$ is the current-current vertex. $G^0$, and $G$ correspond to the Green's function without and with self-energy correction, respectively. The corresponding diagrams are given in Fig.~\ref{Feynmann_diagram}.

It is now easy to deduce that the MT and AL terms scale as $V/\mathcal{W}^4$ and $V^2/\mathcal{W}^6$ where $V$ is the fluctuation potential which scales as $U^2/\mathcal{W}$. Therefore, as long as coupling strength $U\le \mathcal{W}$ these terms have negligible contributions, except near the critical region where either $V$ and/or the Green's function has a singular contribution. Since we are far away from any singular behavior, and we work in the intermediate coupling regime, we can neglect these high order terms.

Finally, using the spectral representation of the Green's function and performing the Matsubara frequency summation as in Eqs.~\eqref{GtoA}-\eqref{afterMatSum}, we arrive at a similar equation for the Kubo term
\begin{eqnarray}
\mathcal{K}^{\rm Kubo}({\bf q},\epsilon_m) &=&\dfrac{C}{3N}\sum\limits_{{\bf k}}\int\limits_{-\infty}^{\infty}\int\limits_{-\infty}^{\infty} \dfrac{d\omega_{1}}{2\pi}\dfrac{d\omega_{2}}{2\pi} \nonumber\\
&&\times A({\bf k},\omega_{1})A({\bf k}+{\bf q},\omega_{2}){\bf v}({\bf k})\cdot{\bf \Gamma}(k,k+q)\nonumber\\
&&\times \dfrac{f(\omega_{1})-f(\omega_{2})}{i\epsilon_{m}-\omega_{2}+\omega_{1}}.
\label{kubofinal}
\end{eqnarray}
Now substituting for the bare current vertex as ${\bf v}={\bf \Gamma}^{(0)}$, and taking the limit of $\epsilon\rightarrow 0$, and ${\bf q}\rightarrow 0$, we obtain Eq.~\eqref{conductivity}.

%
%

\subsection{Vertex correction}\label{Apx:Vertex}

Vertex correction is an important subject in the theories of strong correlation physics. Owing to the conservation laws, there always arise both density-density and current-current vertices in a homogeneous system. One often denotes both by the same symbol $\Gamma$, where a vector symbol ${\bf \Gamma}$ is used for the current vertex, and a scalar symbol $\Gamma$ is used for the density vertex. In the present bubble diagrams for both density-density correlation functions $\chi$, as well as current-current correlation function $\sigma$, the relevant vertex corrections are the three-point vertex functions, as shown by Bethe and Salpeter.\cite{BetheSalpeter} Thanks to the conservation laws, the density and current vertices are related to each other, as shown by Ward, and their relation is known as the Ward identity.\cite{Ward}

In the following descriptions, we use four-component vertex ${{\it \Gamma}}$ which encode the density and current vertices as $({\bf \Gamma},\Gamma)$. The Bethe-Salpeter vertex correction\cite{BetheSalpeter} is written by the self-consistent equations (see Fig.~\ref{fig2a} for the relevant diagram)\cite{TakadaVertex}:
\begin{subequations}
\begin{eqnarray}
\label{BSvertex}
{\it \Gamma}_{\nu}(k,k+q) &=& {\it \Gamma}^{(0)}(k,k+q) + {\it \Gamma}_{\nu}^{(1)}(k,k+q),\nonumber\\
\\
\label{BSvertexb}
{\it \Gamma}_{\nu}^{(1)}(k,k+q) &=& \frac{1}{\beta}\sum_{k',q'}V_{\nu}(k,k+q,k',k'+q) \nonumber\\
&&\times G(k')G(k'+q'){\it \Gamma}_{\nu}(k',k'+q),\nonumber\\
\end{eqnarray}
\end{subequations}
where $\nu=1,2$ are for spin and charge components, respectively. ${\it \Gamma}^{(0)}(k,k+q)$ is the four-component bare vertex, whose density component is $\Gamma^{(0)}=1$. The current components are obtained as ${\bf q}\cdot {\bm \Gamma}^{(0)}=\xi_{k+q}-\xi_k$, where $\xi_k$ is the bare electronic dispersion. ${\it \Gamma}_{\nu}^{(1)}(k,k+q)$ is the first order correction (see Fig.~\ref{fig2a}) to be evaluated self-consistently. Since both spin and charge densities are conserved here, one obtains the same Ward identity for them as
\begin{eqnarray}
&&i\epsilon_m \Gamma_{\nu}(k,k+q) - {\bf q}\cdot {\bf \Gamma}_{\nu}(k,k+q) \nonumber\\
&&\hspace{3cm} = G^{-1}(k+q)-G^{-1}(k).
\label{ward}
\end{eqnarray}
We note that in both Eqs.~\eqref{BSvertexb}, ~\eqref{ward}, the Green's function $G(k)$ is the full self-energy dressed Green's function, which remain the same in both spin and charge sectors. The current vertex does not directly contribute to the density-density correlation, and it is self-consistently related to the current vertex by the Ward identity. Therefore, in an ideal case, one needs to  solve Eqs.~\eqref{BSvertex}, \eqref{BSvertexb}, \eqref{ward} inside the self-consistent cycles for the self-energy calculation. 

Since vertex corrections often make the calculations computationally unmanageable, approximations are inevitable. The zeroth order rule is to make sure the the sum rule is maintained. However, the choice of a given approximation is usually determined by the type of fluctuations one is interested in as well as its region of validity. The simplest one is to neglect the vertex correction. Such an approximation is good enough for electron-phonon coupling (Midgal's theorem),\cite{def_sus} or in the single-shot $GW$ method for electron-electron interactions. Omission of vertex correction can lead to violation of sum rule(s) when self-consistency is invoked.\cite{Kontani_FLEX,DasAIP} The next level approximation is to assume that the density and current vertices are proportional to each other, i.e., $\Gamma_{\nu}={\bf q}\cdot{\bf \Gamma}_{\nu}/i\epsilon_m$ at all momenta and frequencies.\cite{def_sus} Such an approximation yields good result when the momentum dependence of the self-energy is weak as often used in DMFT calculations. However, this can lead to problems when the momentum dependence is significant, simply because the current vertex arises mainly from the momentum derivative of the self-energy.\cite{def_sus} A momentum and frequency dependent ratio function between the density and current vertices was introduced in the literature for the particle-hole bubble interactions\cite{Altshuler,TakadaVertex} as $\Gamma_{\nu}(k,k+q)={\bf B}_{\nu}(k,k+q)\cdot{\bf \Gamma}_{\nu}(k,k+q)$. ${\bf B}={\bf q}/i\epsilon_m$ in the above approximation. Altshuler, {\it et al.}\cite{Altshuler} assumed that the current vertex along the dimension of motion is proportional to the density vertex, which means they ignored multiple scattering channels along the direction of the applied voltage. Takada\cite{TakadaVertex} used the full ratio function, but assumed a local approximation for the potential $V$ ($V(q)$ was replaced by its momentum averaged value), which is again suitable for weak $k-$dependent self-energy.

Eqs.~\eqref{BSvertex}, \eqref{BSvertexb} are required to be solved for either the density or the current term, and then the other term can be evaluated by using the Ward identity (Eq.~\eqref{ward}). This is in fact the best strategy which guarantees that the conservation laws remain intact no matter what approximation is invoked in the calculations. We calculate the current vertex explicitly, and obtain the density vertex from the Ward identity.

For the susceptibility calculation, we assumed a local-field approximation. Therefore, we can make the same local-field approximation for the fluctuation-exchange potential $V$, i.e., we assume $V_{\nu}(k+q,k',k'+q)=V_{\nu}(k+q)\delta_{k,k'}$ (note that we invoked a local filed approximation for both the momentum and frequency axes). Such an approximation should be relaxed when Umklapp scattering or any translational symmetry breaking field is present. From Eq.~\eqref{Eq:Se1}, we can write  $V_{\nu}(k+q)=\frac{1}{\Gamma_{\nu}(k,k+q)}\frac{\delta \Sigma_{\nu}(k)}{\delta G(k+q)}$. Substituting this in Eq.~\eqref{BSvertexb}, we can write,
\begin{eqnarray}
{\bf \Gamma}_{\nu}^{(1)}(k,k+q) \approx  G(k)\Sigma(k)\frac{{\bf \Gamma}_{\nu}(k,k+q)}{\Gamma_{\nu}(k,k+q)}. 
\label{BSvertexb1}
\end{eqnarray}
We define a function $u_{\nu}(k,k+q) = G(k+q)\Sigma_{\nu}(k)$. Then substituting Eq.~\eqref{BSvertex}, we get
\begin{subequations}
\begin{eqnarray}
&{\bf \Gamma}_{\nu}^{(1)}(k,k+q) \approx  \frac{u_{\nu}(k,k+q)/\Gamma_{\nu}(k,k+q)}{1-u_{\nu}(k,k+q)/\Gamma_{\nu}(k,k+q)}{\bf \Gamma}^{(0)}(k,k+q).\qquad\\
\label{BSvertex2}
&{\bf \Gamma}_{\nu}(k,k+q) \approx   \frac{1}{1-u_{\nu}(k,k+q)/\Gamma_{\nu}(k,k+q)}{\bf \Gamma}^{(0)}(k,k+q).\qquad 
\label{BSvertexb2}
\end{eqnarray}
\end{subequations}
Substituting Green's function $G^{-1}(k)=i\omega_n-\xi_{\bf k}-\Sigma(k)$, in the Ward identity in Eq.~\eqref{ward}, we obtain
\begin{eqnarray}
\Gamma_{\nu}(k,k+q) &=& 1- \frac{\Sigma(k+q)-\Sigma(k)}{i\epsilon_m}+\frac{{\bf q}\cdot{\bf \Gamma}_{\nu}^{(1)}(k,p+q)}{i\epsilon_m}.\nonumber\\
\label{Densityvertex}
\end{eqnarray}
We define two symbols $m^*(k,k+q)/m_0 = 1- (\Sigma(k+q)-\Sigma(k))/{i\epsilon_m}$, and $v(k,k+q)={\bf q}\cdot{\bf \Gamma}^{(0)}(k,p+q)/{i\epsilon_m}$. Then substituting Eq.~\eqref{BSvertex2} in Eq.~\eqref{Densityvertex}, we get
\begin{eqnarray}
\Gamma &=& \frac{m^*}{m_0}+v\frac{u/\Gamma}{1-u/\Gamma},
\label{Densityvertex2}
\end{eqnarray}
where we have kept the $k$, and $\nu$ dependence on each term, except $m_0$, implicit, for simplicity. Eq.~\eqref{Densityvertex2} is an algebric equation which can be solved to get
\begin{eqnarray}
\Gamma &=& \frac{m^*/m_0+u\pm\sqrt{(m^*/m_0-u)^2+4uv}}{2}.\nonumber\\
\label{Densityvertex3}
\end{eqnarray}
Eq.~\eqref{Densityvertex3}, and \eqref{ward} can be solved in each self-consistent cycles to obtain both density and current vertices. 

If the self-energy is linear in frequency (FL-ansatz), and linear in momentum, we can further approximate the vertex corrections. Here we get 
\begin{eqnarray}
\frac{m_{\nu}(k,k+q)^*}{m_0}\approx Z^{-1}(k+q)-\frac{{\bf q}}{i\epsilon_m}\cdot\nabla_{\bf k} \Sigma(k),
\end{eqnarray}
and
\begin{eqnarray}
{\bf \Gamma}_{\nu}^{(1)}(k,k+q) \approx \nabla_{\bf k} \Sigma(k).
\end{eqnarray}
This reduces the density and current vertices as\cite{def_sus}
\begin{subequations}
\begin{eqnarray}
\Gamma_{\nu}(k,k+q) &\approx&  m_{\nu}^*/m_0=Z_{\nu}^{-1}(k+q),\\
\label{Densityvertexfinal}
{\bf \Gamma}_{\nu}^{(1)}(k,k+q) &\approx& {\bf \Gamma}_{\nu}^{(0)}(k,k+q) +  \nabla_{\bf k} \Sigma(k)\nonumber\\
&\approx& -m_0{\bm \nabla}G^{-1}({{\bf k}},\omega).
\end{eqnarray}
\end{subequations}

\end{appendix}


\begin{thebibliography}{1}
\bibitem{footnote_tau} In the FL ansatz, electronic scattering is assumed to be solely responsible for transport, and one often gets $\tau^{-1}=-2\Sigma^{\prime\prime}$.\cite{Allen,Markiewicz} This assumption is valid if the system obeys causality, but breaks down if the self-energy is non-analytic. In our case, we do not get any non-analytic self-energy even in the NFL state.
\end{thebibliography}
\end{document}